\documentclass[lettersize,journal]{IEEEtran}

\usepackage[table,xcdraw]{xcolor}
\usepackage{soul}
\usepackage{color}
\usepackage[acronym,shortcuts]{glossaries}
\usepackage{graphicx}
\usepackage{subfigure}
\usepackage{multirow}
\usepackage{balance}
\usepackage{xcolor}
\usepackage{amsmath,amsfonts}
\usepackage{algorithmic}
\usepackage{algorithm}
\usepackage{nicematrix}
\usepackage{makecell}
\usepackage{array}
\usepackage{textcomp}
\usepackage{stfloats}
\usepackage{url}
\usepackage{verbatim}
\usepackage{diagbox}
\usepackage{breakurl}
\usepackage{tikz}
\usepackage{adjustbox}
\usepackage{hhline}
\usepackage{url}

\def\check{\tikz\fill[scale=0.4](0,.45) -- (.35,0) -- (1,.7) -- (.35,.25) -- cycle;}
\def\checkmark{\tikz\fill[scale=0.4](0,.35) -- (.25,0) -- (1,.7) -- (.25,.15) -- cycle;}

\newcolumntype{R}[2]{%
	>{\adjustbox{angle=#1,lap=\width-(#2)}\bgroup}%
	l%
	<{\egroup}%
}
\newcommand*\rot{\rotatebox{ 90}}

\newacronym{ac}{AC}{Alternate Current}
\newacronym{bms}{BMS}{Battery Management System}
\newacronym{can}{CAN}{Controller Area Network}
\newacronym{cps}{CPS}{Cyber-Physical System}
\newacronym{dc}{DC}{Direct Current}
\newacronym{ecu}{ECU}{Electronic Control Unit}
\newacronym{ev}{EV}{Electric Vehicle}
\newacronym{evse}{EVSE}{Electric Vehicle Supply Equipment}
\newacronym{fpga}{FPGA}{Field Programmable Gate Array}
\newacronym{hvac}{HVAC}{Heating, Ventilation, and Air Conditioning}
\newacronym{lin}{LIN}{Local Interconnect Network}
\newacronym{mitm}{MitM}{Man-in-the-Middle}
\newacronym{most}{MOST}{Media Oriented Systems Transport}
\newacronym{tcp}{TCP}{Transmission Control Protocol}
\newacronym{tls}{TLS}{Transport Layer Security}
\newacronym{v2g}{V2G}{Vehicle-to-Grid}
\newacronym{wipt}{WIPT}{Wireless Information and Power Transfer}
\newacronym{wpt}{WPT}{Wireless Power Transfer}
\newacronym{dos}{DoS}{Denial of Service}
\newacronym{soc}{SoC}{State of Charge}
\newacronym{ccs}{CCS}{Combined Charging System}
\newacronym{ocpp}{OCPP}{Open Charge Point Protocol}
\newacronym{plc}{PLC}{Power Line Communication}
\newacronym{pwm}{PWM}{Pulse-Width Modulation}
\newacronym{hil}{HIL}{Hardware-In-the-Loop}

\newacronym{ota}{OTA}{Over-the-Air}
\newacronym{hlc}{HLC}{High-Level Communication}

\newcommand{\parag}[1]{\noindent\textbf{#1. }}

\usepackage{fontawesome}
\newcommand{\bms}{\faBatteryThreeQuarters~}
\newcommand{\controller}{\faCar~} %\faGears }
\newcommand{\wired}{\faPlug~} %\faBolt }
\newcommand{\pad}{\hspace{12pt}}

\newcommand{\wpt}{\faWifi}
\begin{document}

\title{Electric Vehicles  Security and Privacy: \\ Challenges, Solutions, and Future
Needs}

\author{    Alessandro~Brighente,~\IEEEmembership{Member,~IEEE,}~Mauro~Conti,~\IEEEmembership{Fellow,~IEEE,}~Denis~Donadel,~\IEEEmembership{} \\ Raadha~Poovendran,~\IEEEmembership{Fellow,~IEEE,}       ~Federico~Turrin,~\IEEEmembership{}~and~Jianying~Zhou,~\IEEEmembership{Senior Member,~IEEE,}
\thanks{A. Brighente, M. Conti, D. Donadel, and F. Turrin are with the Department of Mathematics and HIT Research Center, University of Padova, 35131 Padova, Italy (e-mail: alessandro.brighente@unipd.it; conti@math.unipd.it;).}%
\thanks{R. Poovendran is with the Department of Electrical and Computer Engineering, University of Washington, 98195 Seattle, USA.}%
\thanks{J. Zhou is  School of Information Systems Technology and Design, Singapore University of Technology and Design, 487372, Singapore. (email : jianying zhou@sutd.edu.sg).}%
\thanks{Manuscript received X X, 2022; revised X X, 2022.}}

% The paper headers
%\markboth{IEEE Transactions on Intelligent Transportation Systems}%
%{Shell \MakeLowercase{\textit{et al.}}: A Sample Article Using IEEEtran.cls for IEEE Journals}

% make the title area
\maketitle

% As a general rule, do not put math, special symbols or citations
% in the abstract or keywords.
\begin{abstract}
Electric Vehicles (EVs) share common technologies with classical fossil-fueled cars, but they also employ novel technologies and components (e.g., Charging System and Battery Management System) that create an unexplored attack surface for malicious users. Although multiple contributions in the literature explored cybersecurity aspects of particular components of the EV ecosystem (e.g., charging infrastructure), there is still no contribution to the holistic cybersecurity of EVs and their related technologies from a cyber-physical system perspective.

In this paper, we provide the first in-depth study of the security and privacy threats associated with the EVs ecosystem. We analyze the threats associated with both the EV and the different charging solutions. Focusing on the Cyber-Physical Systems (CPS) paradigm, we provide a detailed analysis of all the processes that an attacker might exploit to affect the security and privacy of both drivers and the infrastructure. To address the highlighted threats, we present possible solutions that might be implemented. We also provide an overview of possible future directions to guarantee the security and privacy of the EVs ecosystem. Based on our analysis, we stress the need for EV-specific cybersecurity solutions.

\end{abstract}


\IEEEpeerreviewmaketitle

\glsresetall
\section{Introduction}

\IEEEPARstart{T}{he} recent climate crisis demands green alternatives to replace technologies with high environmental impact. Among the others, fossil-fueled transportation is one of the significant causes of greenhouse gases. \acp{ev} have been proposed as a green alternative, where electric batteries are employed as a power source. During the last years, the number of people opting for the \ac{ev} alternative increased up to the point where the market share of new \ac{ev} sales reached more than $50\%$ in countries such as Iceland ($55.6\%$) and Norway ($82.7\%$)~\cite{evShare}. The adoption of \acp{ev} is further expected to increase in the next years. In fact, governments are incentivizing the adoption of \acp{ev} thanks to the deployment of a large number of \ac{evse} in public charging infrastructures~\cite{biden} and planning to ban sales of fossil-fueled vehicles~\cite{nether}. Furthermore, technology advancements remove the current barriers against consumers' adoption of \acp{ev}, providing extended driving range and seamless charging~\cite{capuder2020review}.

The increasing number of \acp{ev} demands a thorough analysis of the security of both vehicles and infrastructure operations. 
%In fact, \acp{ev} are equipped with a large number of \acp{ecu} that measure information regarding different parts of the \ac{ev} and its status %\denis{ok, but also fuel vehicles have a lot of ECU..}. 
%Each \ac{ev} is equipped with actuators that control how the vehicle responds to the different inputs both from inside and outside the vehicle. 
Like traditional vehicles, \acp{ev} are equipped with many \acp{ecu}, sensors and actuators that measure, process, and control the different stimuli inside and outside the vehicle. 
However, \acp{ev} include additional components. Indeed, an \ac{ev}  integrates components to govern the hardware and software dedicated to managing electric energy smartly. These components are, for instance, the \ac{bms} and the charging system.

\begin{figure}[h]
    \centering
    \includegraphics[width=\columnwidth]{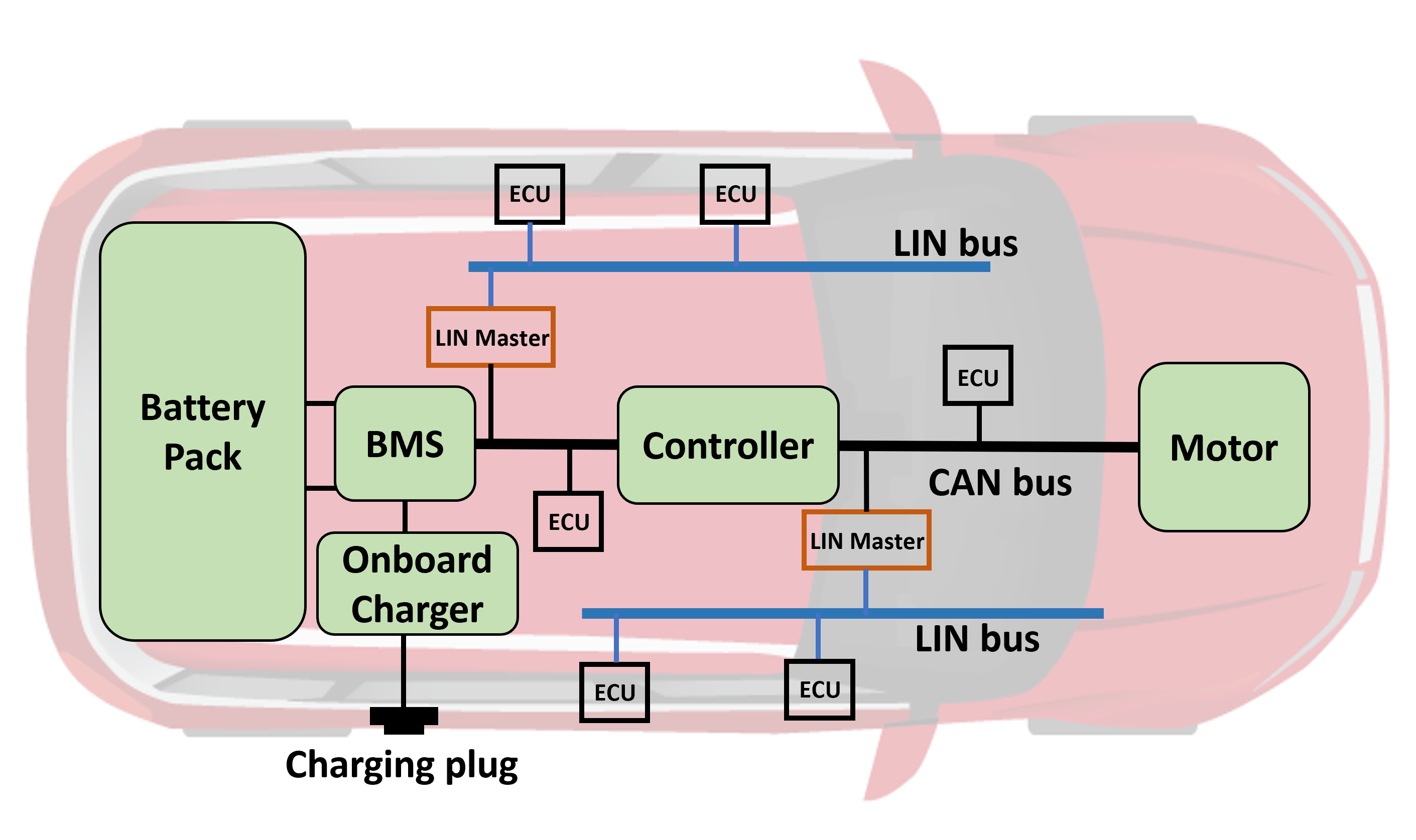}
    \caption{Main components of an \ac{ev}. Green components are \ac{ev} specific.}
    \label{fig:components}
\end{figure}

Different studies have already proven the impact of potential cybersecurity attacks on automotive systems. For instance, Miller and Valasek~\cite{miller2015remote} proved the feasibility of hijacking a vehicle by remotely controlling it through the infotainment system. Furthermore, most existing vehicles exploit \ac{can} as in-vehicle network architecture, which has already been proved as non-secure~\cite {lin2012cyber} and, therefore, may be vulnerable to potential cyberattacks. Lastly, privacy shall also be guaranteed to prevent malicious users from obtaining sensitive information on the driver, such as her location or habits. It is essential to include security and privacy features by design to prevent these and other attacks. Researchers investigated vehicle security, focusing on the different aspects of in-car communications~\cite{schmittner2019automotive, scalas2019automotive}. However, \acp{ev} are equipped with specific components that provide fundamentally different attack surfaces and exploitation points. For instance, \acp{ev} are equipped with electric batteries to power the vehicle components ranging from the infotainment system to the acceleration pedal, together with systems to manage the electrical power as shown in Figure~\ref{fig:components}. 

Therefore, analyzing the in-vehicle threats associated with these components is essential. Furthermore, the power supply must be regulated by dedicated hardware, not classical vehicles. Researchers discussed how the \ac{ev} charging infrastructure could be exploited by attackers~\cite{gottumukkala2019cyber}, however, neglecting the in-vehicle threats.

\parag{Contribution}
In this paper, we examine the security and privacy issues of \acp{ev} from a \ac{cps} point of view. Given the high demand for \acp{ev} and the increasing number of deployed charging facilities, it is fundamental to guarantee the security and privacy of both vehicles and users.
Many literature contributions discuss solely technical aspects of the \ac{ev} ecosystem without focusing on security issues. Other security-focused works study a single system component (e.g., the vehicle's internal bus, the smart grid, or the communication protocols) without comprehensively analyzing the whole environment.
% Furthermore, researchers approaching \ac{ev} security cannot reach those contributions focusing on specific components, as they may lack the needed knowledge on their properties and functioning. 
% Towards this, 
We provide a general overview of \ac{ev} functioning, focusing on their core components to build the basic knowledge needed to analyze the possible threat vectors.
We then discuss possible attacks and countermeasures specific for \ac{ev} and underline the existing security solutions for fuel vehicles that are also effective in \acp{ev}. 
% In this work, we focus on systems which take advantage of peculiar features generated by the presence of electric components}. \fede{Non convinto dell'ultima frase in questo punto}\ale{in particolare? modifica proposta?}
With the bird-eye on the \ac{cps} concept, we are not only able to discuss the issues related to the exchange of information between the different involved entities, but also the side channels that may leak sensitive information or that could lead to hazardous behavior impacting on users' safety.
We hence shed light on the unresolved challenges of \acp{ev} ecosystems, providing interested researchers with possible directions worth investigating to guarantee the security and privacy of the overall \ac{ev} ecosystem. 
We also consider future directions such as the \ac{wpt} charging of \acp{ev}, which has only been developed on small-scale testbeds at the time of writing. 
We believe that delving into this emerging system's security and privacy issues will help future developers design and implement secure-by-design \ac{wpt} solutions for \ac{ev}.

We summarize the contribution of this work as follows.
\begin{itemize}
    \item We examine the peculiar components that differentiate \acp{ev} from fossil-fueled vehicles and provide an overview of their role and how they exchange information. 
    \item We provide an overview of the different technologies employed to charge electric vehicles, comprising both wired and wireless charging. We present the available standards for each of them and describe their basic functioning.
    \item We provide an in-depth discussion of the security and privacy issues of the \acp{ev} ecosystem. We analyze the threats related to the in-vehicle network and the threats related to the charging process. We particularly focus on their effects on the peculiar \ac{ev} components and analyze them from a \ac{cps} point of view.
    \item We analyze and compare possible countermeasures proposed in the literature for each of the presented attacks, even grasping from other similar areas.
    \item We outline future directions for research in the \ac{ev} cybersecurity domain.
\end{itemize}

\parag{Organization}
The rest of the paper is organized as follows. In Section~\ref{sec:relLit}, we review the related literature. In Section~\ref{sec:evFrom}, we describe the \ac{ev} components the charging infrastructure. In Section~\ref{sec:sepChallenges}, we then discuss the in-vehicle security and privacy threats, and those related to the charging infrastructure in Section~\ref{sec:charginSep}. Along with the threats, we also present possible countermeasures. Then, we discuss the possible future direction in Section~\ref{sec:futureDir}, and lastly we conclude the paper in Section~\ref{sec:conc}.

\section{Related Literature on EV Security}\label{sec:relLit}

Automotive cyber-security requires standardization to allow for security guarantees and interoperability. Schmittner et al.~\cite{schmittner2019automotive} reviewed the available standards, including designing and validation aspects. These standards, however, do not consider the peculiar features of \acp{ev}. Scalas et al.~\cite{scalas2019automotive} provided an overview of the cybersecurity requirements for the future of the automotive industry, focusing on in-vehicle components. They discussed several technologies and attacks but were not specific for \acp{ev}.
Furthermore, different works present technical reviews of the \ac{ev} ecosystem~\cite{bahrami2020ev, das2020electric}. However, none of them consider the security aspects.

Some contributions in the literature focused on specific components of \acp{ev}. For instance, Khalid et al.~\cite{khalid2019facts} focus on the \ac{bms}, discussing the lack of a cybersecurity standard to guarantee its security and providing an overview of the possible standardization framework that could be adopted to achieve this goal. Chandwani et al.~\cite{chandwani2020cybersecurity} presented an overview of the cybersecurity threats associated with the onboard charging system of \acp{ev}. Despite providing an accurate analysis of the security of this component, their contribution does not consider how these attacks can impact the other peculiar components of the \ac{ev}. These contributions do not provide a general overview of the \ac{ev} ecosystem. Furthermore, they do not discuss the threats associated with privacy. 
Acharya et al.~\cite{acharya2020cybersecurity} provide the first discussion on how \acp{ev} can be considered as \ac{cps}. The authors discuss how different attacks can be conducted inside the car and during communication with the power supplier. However, they do not consider the specific components of the \acp{ev} such as the \ac{bms}. Jin et al.~\cite{ye2020cyber} focus on the \ac{cps} system represented by the power electronics in \acp{ev}. However, they do not consider how these attacks may impact the other components of the \ac{ev} and did not discuss the issues related to \ac{wpt}. Most of the literature related to \acp{ev}' cybersecurity focus on the charging infrastructure and process. Gottumukkala et al.~\cite{gottumukkala2019cyber} provide an overview of the \ac{cps} threats associated with a wired \ac{ev} charging infrastructure. Antoun et al.~\cite{antoun2020detailed} discuss the security threats associated with the negotiation and actuation of a charging session investigating the communications between the multiple involved entities. They presented different charging scenarios, neglecting the \ac{wpt} option. 

{\renewcommand{\arraystretch}{1.2}
\begin{table*}[!h]
\label{tab:litRev}
\centering
\resizebox{\textwidth}{!}{%
\begin{tabular}{|c|c|c|c|c|c|c|c|c|}
        \hline
        \textbf{Reference} & \textbf{\ac{bms}} & \textbf{Onboard Charger} & \textbf{Battery Pack} & \textbf{Controller} & \textbf{Electric Motor} & \textbf{Wired Charging} & \textbf{Wireless Charging} \\ \hline %\hhline{|=|=|=|=|=|=|=|=|}
        Khalid et al. \cite{khalid2019facts} & \checkmark &  &  &  &  &  &  \\ \hline
        Chandwani et al. \cite{chandwani2020cybersecurity} &  & \checkmark &  &  &  &  &  \\ \hline
        Acharya et al. \cite{acharya2020cybersecurity} &  &  &  &  &  & \checkmark &  \\ \hline
        Ye et al. \cite{ye2020cyber} & \checkmark &  &  & \checkmark & \checkmark & \checkmark &  \\ \hline
        Sripad et al. \cite{sripad2019Vulnerabilities} &  &  & \checkmark &  &  &  &  \\ \hline
        Gottumukkala et al. \cite{gottumukkala2019cyber} &  &  &  &  &  & \checkmark &  \\ \hline
        Antoun et al. \cite{antoun2020detailed} &  &  &  &  &  & \checkmark &  \\ \hline
        %Fraiji et al. \cite{fraiji2018cyber} & \checkmark &  &  &  &  &  &  \\ \hline
        Garofalaki et al. \cite{garofalaki2022electric} &  &  &  &  &  & \checkmark &  \\ \hline
        Van Auben et al. \cite{van2022security} &  &  &  &  &  & \checkmark &  \\ \hline 
        Babu et al. \cite{babu2022survey} &  &  &  &  &  &  & \checkmark \\ \hline 
        \textbf{Our paper} & \check & \check & \check & \check & \check & \check & \check \\
        \hline
    \end{tabular}}
\end{table*}
} % closing array stretch

Vehicles can be interconnected with one another to form the internet of vehicles. This is also feasible with \acp{ev}, which imposes additional security challenges. Fraiji et al.~\cite{fraiji2018cyber} discuss the cybersecurity threats associated with the internet of electric vehicles, discussing the threats associated with the communication with the multiple involved entities being part of the road infrastructure. The cybersecurity focus is on the communication links, therefore neglecting the impact of the peculiar \acp{ev}' components. 

Garofalaki et al.~\cite{garofalaki2022electric} present a detailed survey on \ac{ocpp} and the corresponding threat and vulnerabilities on the \ac{v2g} ecosystem due to its adoption. Similarly~\cite{van2022security} overview the main protocols for~\ac{ev} charging adopted in the Netherlands and analyze their security features, while Babu et al.~\cite{babu2022survey} analyzed the security of the main protocols proposed for the~\ac{ev} environment with a particular focus on the payment methods and the authentication solutions. 
Differently from these works, instead of focusing on the protocols, we focus on the entire~\ac{ev} architecture, highlighting the main security and privacy challenges in this typology of \ac{cps}.

Table~\ref{tab:litRev} compares the related works on \ac{ev} security with our contribution. We can see that most of the contributions focus on vehicle-to-grid communications in the wired case. However, none of the available papers focus on intertwining the different cyber-physical aspects of \acp{ev}. Therefore, our paper provides a more complete analysis of the security and privacy challenges for \acp{ev}.

\section{Electric Vehicles from a Cyber-Physical System Perspective}\label{sec:evFrom}

In this section, we analyze \acp{ev} from a \ac{cps} perspective. We emphasize those components that differentiate \acp{ev} from gas-fueled vehicles. In particular, we first describe the traditional vehicle architecture in Section~\ref{sec:bus}. Then, we present the main components of an \ac{ev} in Section~\ref{sec:components}, showing how it differs from traditional vehicles. Lastly, we provide an overview of the \ac{ev} charging infrastructure in Section~\ref{sec:charging}.

\subsection{Traditional Vehicle Architecture}\label{sec:bus}
% \denis{add vehicle network composition more in detail (not needed)}

Nowadays, vehicles contain dozens of different microcomputers, called \acfp{ecu}, running millions of lines of code~\cite{Many22}. Each \ac{ecu} is responsible for controlling a mechanical (e.g., brakes) or electrical (e.g., light) component of a modern vehicle.
Depending on the component it has to manage, an \ac{ecu} generally employs a wide range of microcontrollers, from simple $8$-bit RISC controllers to more complex $32$-bit multicore processors. \ac{ecu} are typically implemented with ad-hoc firmware, even if complex \acp{ecu} may run complete operating systems: the infotainment system, for instance, usually runs a Linux-based kernel. In order to provide more flexibility during updates, more advanced solutions envisage the implementation of multiple \acp{ecu} on a single FPGA board~\cite{cho2021FPGABased}.

Communications among \acp{ecu} that reside in the vehicle pass through wires that connect multiple components. The two mostly implemented technologies are \ac{can} and \ac{lin}. \ac{can} represents the main network that allows for cost-effective wiring, self-diagnosis and error correction~\cite{bozdal2020evaluation}. The \ac{can} bus consists of two wires and implements a distributed architecture, where car modules (i.e., the \acp{ecu}) share messages upon winning a contention phase. However, \ac{can} has been designed to be a reliable solution, neglecting possible security ad privacy shortcomings. 

The \ac{lin} bus is a supplement to \ac{can}~\cite{ISO17987}. In particular, it connects a smaller number of \acp{ecu} (one master and up to $16$ slave nodes) and offers a drastically cheaper implementation at the cost of lower performance and reliability. A \ac{lin} master node is typically a gateway to \ac{can}, and multiple \ac{lin} buses can communicate via the \ac{can} bus. The \ac{lin} bus can be used to control, among the others, sensors and actuators for steering wheels, comfort, powertrain, engine, air conditioning, doors, and seats.

Besides \ac{can} and \ac{lin}, also other technologies such as FlexRay~\cite{makowitz2006flexray} and \ac{most}~\cite{fijalkowski2011media} are currently used for automotive networks. To overcome some of these technologies' limitations and ease their interoperability, automotive Ethernet has recently been introduced as a possible solution~\cite{corbett2016automotive}. Given the \ac{cps} nature of our investigation, we do not prefer one technology over another, as these all represent communication means for the exchange of information inside the \ac{ev}. We refer the interested reader to~\cite{corbett2016automotive} for a discussion on automotive Ethernet security.

Modern vehicles also include mechanisms to update the internal software. This service is generally implemented with the aid of external device plug (e.g., USB flash drive) or \ac{ota} software update~\cite{halder2020secure} (e.g., via Internet connection).
Furthermore, many vehicles nowadays include complex entertainment systems, which may expand the vulnerable surface, exposing new connections (e.g., Bluetooth) and operating systems (e.g., Android).
% LIN and CAN
% https://www.csselectronics.com/pages/lin-bus-protocol-intro-basics

\subsection{Electric Vehicle Specific Components}\label{sec:components}
\acp{ev} share most of the architecture with fuel-based vehicles. However, they comprise a different set of hardware modules that manage how the vehicle generates power and how to generate motion. In particular, an \ac{ev} comprises the following components~\cite{un2017comprehensive}, depicted in Figure~\ref{fig:components}.

\parag{Battery} The battery is where the charge is stored in the form of \ac{dc}. It provides the power needed to operate the \ac{ev} components. Batteries are usually combined in packs and connected in series or parallel to increase the voltage and Amper/hour they can deliver to the \ac{ev}. Batteries suitably combined are enclosed into a metal casing to prevent damage. The case usually includes a cooling system to avoid damage due to batteries overheating.
    
\parag{Battery Management System} This module manages all operations regarding the battery. It manages the current output and the charging and discharging of the battery by keeping it in a safe operating area. Hence, it regulates the electricity flow through the battery. The \ac{bms} is unique for each \ac{ev} model, and may be designed according to various topologies, i.e., modular, centralized or distributed~\cite{khalid2019facts}. The \ac{bms} monitors each battery in the pack and measures each cell's voltage, current, and temperature. It is instructed with a threshold limit for each of them and disconnects the load if values exceed the threshold value. Furthermore, the \ac{bms} measures the \ac{soc} and state of health of the battery. The \ac{bms} communicates with the human-machine interface to report information on this information. All the information are exchanged via \ac{can} or \ac{lin} bus.
     
\parag{Battery Charger/Onboard Charger} This component provides an interface between the charging system and the \ac{ev} battery. As soon as an \ac{ac} charging process begins, the charger converts the input voltage to \ac{dc} and passes it to the battery for storage. For high power \ac{dc} charging, the conversion phase is done on the charging column. Furthermore, it prevents possible damages to the battery or the supply system (e.g., overheating) by limiting the power flow~\cite{chandwani2020cybersecurity}.
    
\parag{Controller} The controller handles the flow of current from the battery to the \ac{ev} associated with all operations, ranging from motors-related operations to powering the infotainment system. It receives the input from the driver to control the acceleration, brake pressure, and driving mode and converts the energy in the battery from \ac{dc} to \ac{ac} to control the \ac{ev} accordingly. On the other hand, the \ac{ev} may generate electricity due to, e.g., regenerative braking. In this case, the controller converts the generated \ac{ac} to \ac{dc} such that the energy can be stored in the battery.
    
\parag{Electric Motor} The motor is powered by the \ac{ev} battery, which provides the electricity needed to turn it and move the \ac{ev}. The electric motor communicates with sensors and actuators in the \ac{ev} that control the amount of thrust required~\cite{guo2020cyber}. There exist many implementations of electric motors. The most commonly used for \acp{ev} are \ac{ac} induction due to their lower cost implementation thanks to the absence of permanent magnets.

These components characterize an \ac{ev} and differentiate it from other types of vehicles. In particular, the conventional motor is replaced with an electric one, and a battery pack replaces the fuel tank. Notice that all the components mentioned above need to share messages inside and outside the vehicle to guarantee the correct functioning. An attacker might exploit some of these messages to create inconsistencies on the \ac{ev} status or to cause damages to both the vehicle and driver. We provide a detailed security analysis based on these components in Section~\ref{sec:sepChallenges}.

%influence of HVAC on battery

%%%%%%%%%%%%%%%%%%%%% EV CHARGING INFRASTRUCTURE %%%%%%%%%

\subsection{Electric Vehicles Charging Infrastructure}\label{sec:charging}

The \ac{ev} needs to charge its battery periodically to provide power to its components. To this aim, the \ac{ev} shall be connected to a charging infrastructure with whom it negotiates a charging session. According to the negotiated session, the infrastructure then delivers the needed energy to the \ac{ev}. Charging may happen either in public areas (e.g., shopping malls or offices) or at a private site (e.g., home). To prevent possible malfunctioning, the charging infrastructure must be carefully managed. This is particularly true when considering a scenario where handling a massive number of \acp{ev} may lead to blackout and other grid malfunctions~\cite{ravi2022utilization}. 

Charging an \ac{ev} differs from other devices, such as smartphones or laptops, as it requires dedicated hardware and a drastically larger energy supply. Indeed, if many \acp{ev} are concurrently charging, there can be grid overloading, leading to malfunctions and local blackouts. To avoid these issues, the grid must employ a communication channel with the \ac{ev} to negotiate to charge parameters that respect the vehicle's battery requirements without overloading the grid. \ac{v2g} refers to the technology enabling this communication type. There are two solutions to manage a charging session: wired and wireless. While the former is more diffused and widely implemented nowadays, the latter is still in the initial stage and under development. Unfortunately, there is no unique world standard to regulate this communication channel. Instead, different manufacturers implement different standards based on the technologies used for the charging process. For instance, CHAdeMO~\cite{chademov2x} (Japan) or GB/T~\cite{liReplayAttackDefense2019} (China) can be used only with wired charging, while ISO 15118 (Europe, North America) also supports \ac{wpt}~\cite{iso15118-8, iso15118-20}.

\subsubsection{Wired Charging}\label{sec:wired}

With this setting, the \ac{ev} is connected to an \ac{evse} through a cable that transmits both the control signals ad the charging current. 
In turn, \acp{evse} negotiate with power grids for the energy needed to charge the vehicle, based on both \ac{ev} and grid requirements. However, these basic functions are integrated by every charging standard, which employs different communication methods. Low current charging levels, such as \ac{ac} Level~1 or \ac{ac} Level~2, require a simple control channel which is generally provided by a \ac{pwm} communication. More advanced charging, such as \ac{dc} charging, needs better management of the energy provided by a \ac{hlc} provided by protocols such as \ac{can} or \ac{plc}. These technologies enable the development of additional services, such as the automatizing of the billing process~\cite{autocharge, iso15118-1}, or the download of firmware updates~\cite{buschlinger2019plug}.
%\st{Furthermore, they may be equipped with modules to automatically authenticate the user and with meters for automatic billing.} 
In case of a lack of automated authentication solutions, \ac{evse} may be equipped with RFID readers through which users can authenticate and pay for the service. \acp{evse} can be deployed at private or public premises: private charging columns are generally less advanced and support less charging level with respect to public \ac{evse}.

There are mainly two protocols supporting the \ac{hlc}s between \ac{ev} and \ac{evse} during \ac{dc} charging sessions. The first one, employed by \ac{ccs}, is the ISO 15118~\cite{iso15118-1}, which modulates data over the control pilot pin using \ac{plc}.
The second one, CHAdeMO~\cite{chademov2x} employs a \ac{can} channel for the communication.

% \denis{vedere cosa salvare di quanto segue:}
% The digital communication protocol used by the \ac{ev} and \ac{evse} to communicate before and during the charging session is based on the ISO 15118~\cite{iso15118-1}. This protocol assumes that both the \ac{ev} and \ac{evse} are able to encrypt and decrypt messages, such that they are able to establish a secure communication to prevent malicious actors to alter the billing information. However, this feature is exploited only for charging session negotiation and billing purposes. We discuss in Section~\ref{sec:sepChallenges} how this might not be sufficient.

\begin{figure}[!h]
	\subfigure[center][Level 2]{
	\includegraphics[width=0.31\columnwidth]{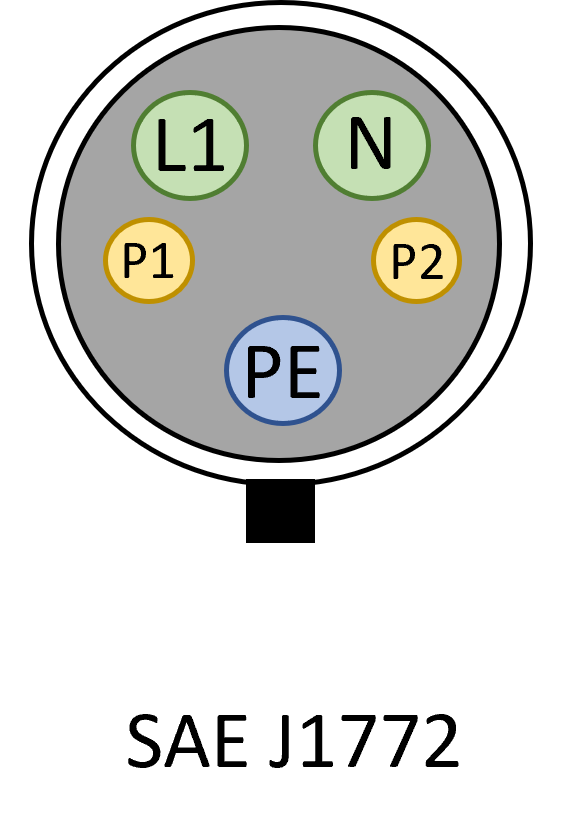}\label{subimg:level2}}
	\subfigure[center][Level 3]{
		\includegraphics[width=0.66\columnwidth]{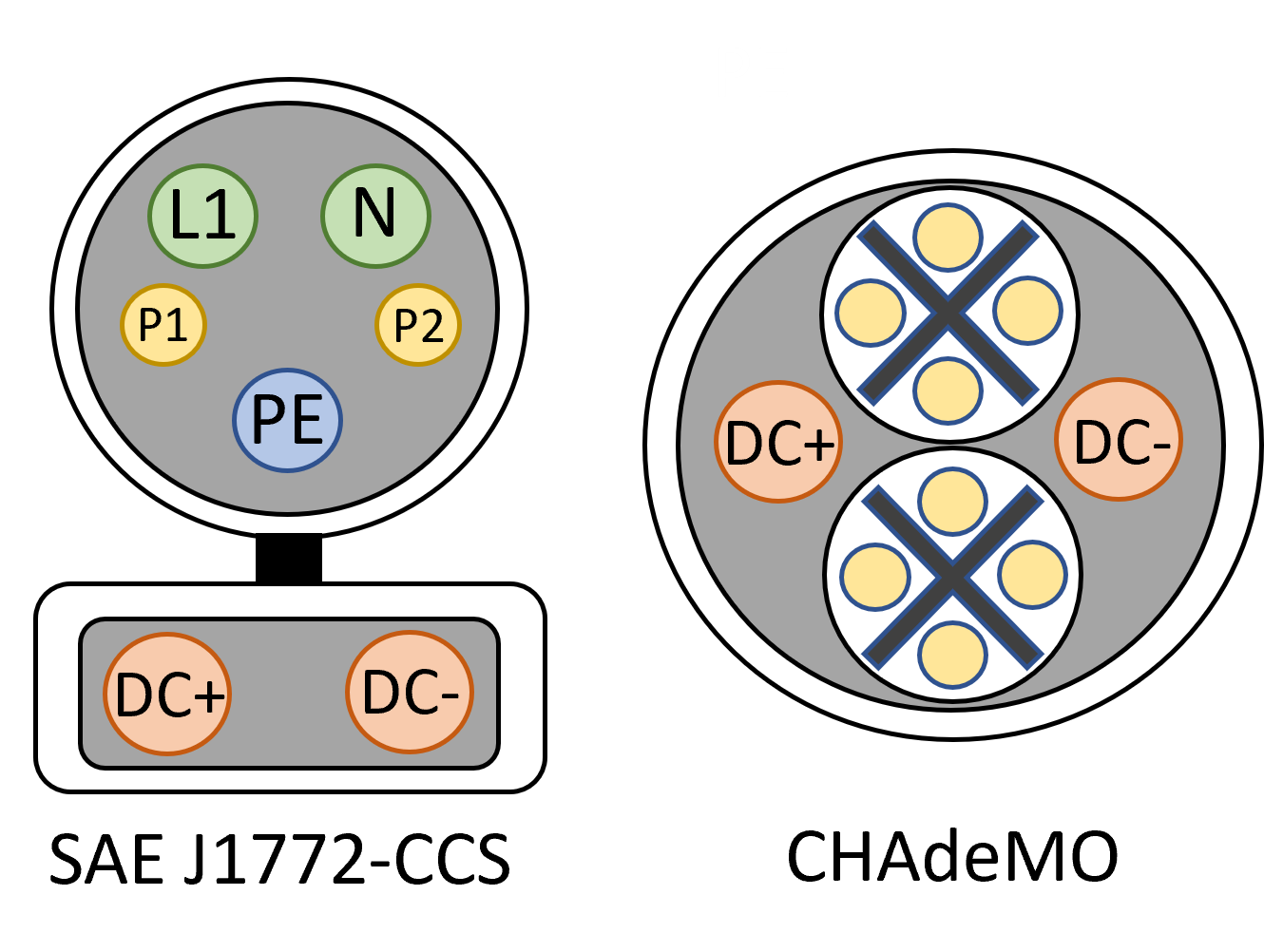}\label{subimg:level3}}
		\caption{Different types of \ac{ev} chargers. L1 = \ac{ac} line 1; N = AC line neutral, P1 and P2 = proximity lines, PE = ground.}\label{fig:chargers}
\end{figure}

The physical connection between \ac{ev} and \ac{evse} may be implemented with different plugs according to different standards. In particular, we can classify \acp{evse} according to different levels. Figure~\ref{fig:chargers} shows the different charger levels together with their lead characterization.

Level~1 and Level~2 \acp{evse} exploit a five-leads connector implementing the SAE J1772 protocol~\cite{SAEJ1772}, as shown in Figure~\ref{subimg:level2}.
This connector exploits two leads to deliver the charging current, two leads for pilot signals, and one lead for ground (or protective earth). The two current leads plus the ground one are used by the \ac{evse} for metering and computing the session cost. The two pilot lines have two different functions. The first one, the control pilot, is used to exchange information with the \ac{ev} during the charging session. The signals exchanged through the control pilot either control the amount of current delivered to the \ac{ev}~\cite{lee2021adaptive} or are used to check the connection status and remove power from the adapter in case of disconnection to prevent the user injuries~\cite{gottumukkala2019cyber}. The second pilot line is the proximity pilot, used by the \ac{ev} to check whether a proper physical connection has been established with the \ac{evse}.

Level~3 \acp{evse}, i.e., those allowing for fast charging, are based on different implementations and are showed in Figure~\ref{subimg:level3}. The first is the \ac{ccs} expansion of the SAE J1772, which allows for direct current exchange for fast charging. Furthermore, it implements \ac{plc} to exchange information between the \ac{ev}, the \ac{evse}, and the smart grid~\cite{SAEJ1772}. The second implementation is the Japanese CHAdeMO~\cite{7400449}, which implements a fast charging protocol. Besides delivering power, this implementation allows for data exchange via the \ac{can} bus protocol. Thanks to this type of connection, it is possible to avoid applying power to the connector in case of a non-safe connection or to exchange information related to the battery \ac{soc}. Furthermore, CHAdeMO allows \ac{v2g} communication, where the \ac{ev} battery is later used as energy storage to provide service to the grid~\cite{chademov2x}. Other protocols exist, such as the proprietary protocol employed by Tesla vehicles and the Chinese GB/T, which will probably be replaced by Chaoji, an evolution of CHAdeMO~\cite{bahrami2020ev}. 

The main differences between Level~3 and Level~2 chargers lie in the higher number of leads in Level~3, and in the implemented circuitry which converts \ac{ac} to \ac{dc}, which is inside the charging columns for Level~3, while it is onboard in the \ac{ev} for Level~2.
Furthermore, Level~3 charging includes richer communication capabilities thanks to the support of \ac{hlc}.

\subsubsection{Wireless Power Transfer}\label{sec:wpt}

Charging via \ac{wpt} allows charging an \ac{ev}'s battery without physically connecting the vehicle to the charging infrastructure. In \ac{wpt}, a source (powered by the grid) generates a time-varying electromagnetic field that triggers the generation of a current at the receiver's (\ac{ev}'s) side. This current is generated thanks to a coil mounted on the \ac{ev}'s side that receives the transmitted electromagnetic field and, due to Faraday's law of induction, generates an \ac{ac}~\cite{zhang2018wireless}. 

\begin{figure}[!h]
    \centering
    \includegraphics[width=\columnwidth]{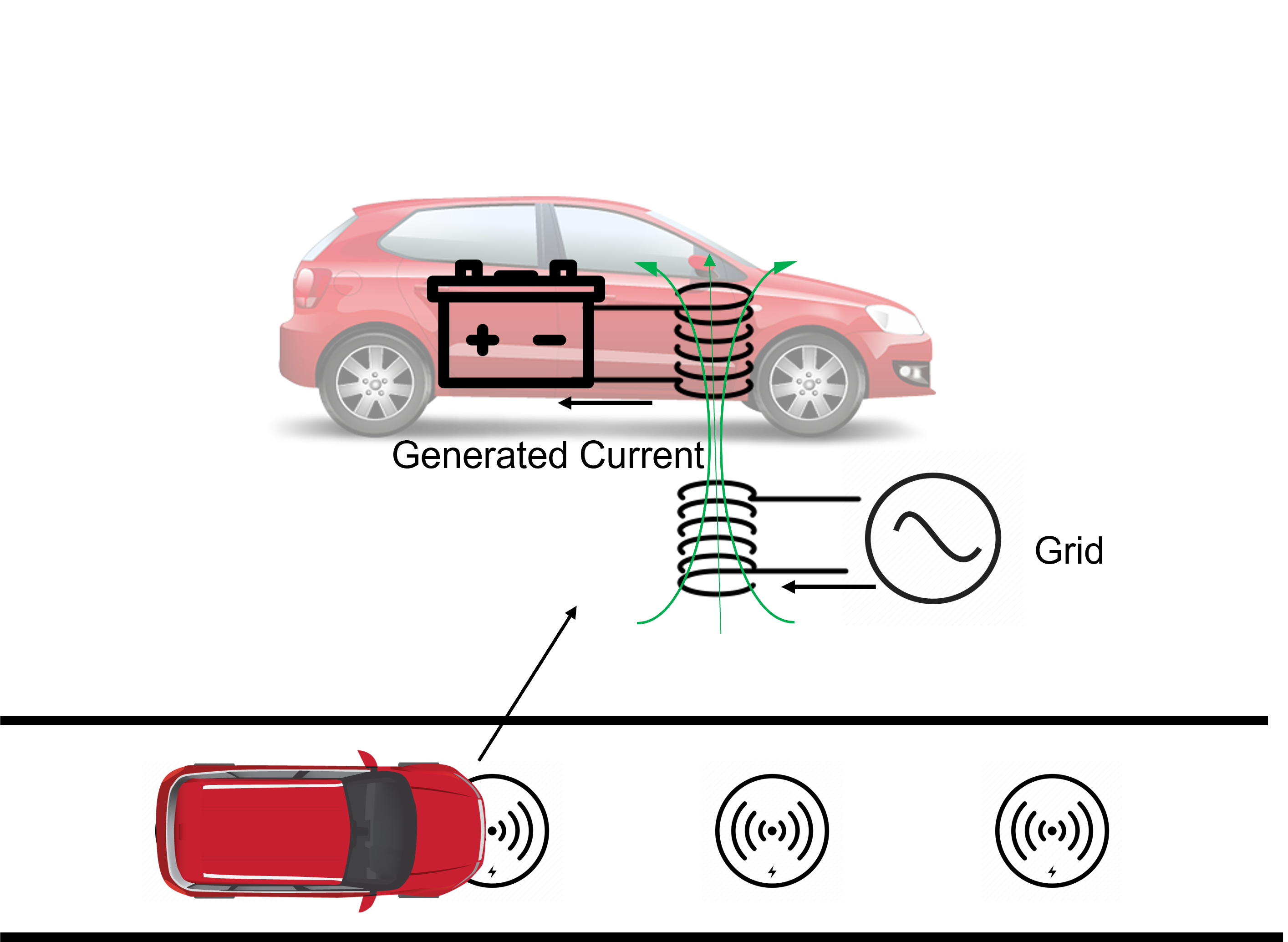}
    \caption{Representation of a \ac{wpt} system for \acp{ev}.}
    \label{fig:wpt}
\end{figure}

Via \ac{wpt}, it is possible to create multiple charging scenarios depending on the mobility of the \ac{ev}~\cite{wang2021improved}. In fact, thanks to the absence of a physical connection, \acp{ev} can be either charged while parked or while driving in a dynamic scenario~\cite{babu2022survey}. The static scenario is similar to the one previously described in Section~\ref{sec:wired}, where a user books a charging session and receives the power from the grid while parked at a charging facility. Instead, the dynamic case requires a suitably designed infrastructure composed of multiple sequential \ac{wpt} transmitters. Figure~\ref{fig:wpt} shows a pictorial representation of a dynamic \ac{wpt} system for \acp{ev}. The street is equipped with multiple \ac{wpt} transmitters deployed underneath the street. These transmitters are connected to the grid that provides the power needed to charge the \ac{ev}.

Dynamic \ac{wpt} can be further divided into two categories: quasi-dynamic and fully-dynamic~\cite{machura2020driving}. In the former case, charging is limited to the cases where the \ac{ev} is not moving, e.g., while waiting at stops or traffic lights. In fully-dynamic \ac{wpt}, charging is continuously delivered to the \ac{ev} as long as it drives near transmitting coils. In both dynamic scenarios, the challenge is to guarantee that transmitters are activated only when needed to avoid energy waste and that only legitimate users access the emitted power. In fact, due to the absence of a medium, users could steal power by driving close to an \ac{ev} that paid for the charging session. We discuss all the security problems related to \ac{wpt} in Section~\ref{sec:sepChallenges}.

As specified in the ISO 15118 standard~\cite{iso15118-8}, the connection between the vehicle and the charging column during a static \ac{wpt} scenario uses WiFi (IEEE 802.11). The vehicle can connect before being correctly parked or when it is already over the coil. If needed, the \ac{evse} provides the \ac{ev} with fine positioning messages to help the driver correctly place the vehicle to reduce energy dispersion. After establishing the connection, the two entities communicate similarly to wired cases. Some modifications are introduced to adapt to the wireless scenario, including the \ac{wpt} charging mode and the fine positioning messages.

Due to their novelty, dynamic and quasi-dynamic charging are not yet covered by approved and widely adopted standards. Some research works~\cite{roman2020authentication} adopt Dedicated Short-Range Communications (DSRC) to create a channel between vehicles and the Road Side Units (RSUs), which are in charge of controlling a portion of the road coils.

Another possible solution can be to extend \ac{wpt} to deliver also information along with power. In fact, \ac{wipt} represents a technology that might be exploited for electric vehicle~\cite{corti2020simultaneous}. \ac{wipt} can be adopted to implement a system similar to that exploited in wired \ac{ev} charging, where control signals are exchanged through the pilot line and the charging current. In \ac{wipt}, control signals can be coded into the time-varying electromagnetic field to deliver power and check the connection's status simultaneously. Furthermore, this solution can be exploited to authenticate \acp{ev} and solve part of the security challenges in \ac{wpt}. Although not yet discussed in the literature, we believe that \ac{wipt} represents a suitable line of research for \ac{ev} charging technologies.

%%%%%%%%%%%%%%%%%%%%%%%%%%%%%%%%%%%%%%%%%%%%%%%%%%%%%%
%%%%%%%%%%%%% IN VEHICLE SEC %%%%%%%%%%%%%%%%%%%%%%%%%

\section{In-Vehicle Security and Privacy Challenges}\label{sec:sepChallenges}

This section discusses the security and privacy challenges related to the components and protocols used inside \acp{ev}. First discuss in Section~\ref{sec:battAndBMS} the challenges related to the battery and the \ac{bms}. Then, we discuss the challenges related to the controller and charger in Section~\ref{sec:cnotrollerandcharger}. The security of \ac{can} bus has been extensively studied in the literature, as it does not envision secure by design solutions~\cite{jo2021survey}. However, how these attacks may impact \acp{ev} has never been studied. Since all in-vehicle messages are exchanged through \ac{can} and \ac{lin} buses, we discuss how their vulnerabilities can be exploited to impact those components specific to the \acp{ev}. We summarize in Table~\ref{tab:inVeh} the in-vehicle security and privacy challenges together with their effects, impact severeness, and possible countermeasures.

{\renewcommand{\arraystretch}{1.2}
\begin{table*}[t]
\label{tab:inVeh}
\caption{Summary of In-Vehicle Challenges.}
\centering
\resizebox{\textwidth}{!}{%
\begin{tabular}{|c|l|l|l|l|}
\hline
\multicolumn{1}{|l|}{\textbf{Component}} &
  \textbf{Attack Type} &
  \textbf{Effect} &
  \textbf{Impact} &
  \textbf{Possible Solutions} \\ \hline
 &
  \acrshort{dos} &
  \begin{tabular}[c]{@{}l@{}}Prevent energy delivery \\ Prevent information reception\\ Increase energy consumption\\ Physically damage the battery\end{tabular} &
  Medium &
  \begin{tabular}[c]{@{}l@{}}Flow control\\ Time-lock puzzles\\ Rate limiting\\Intrusion detection\end{tabular} \\ \cline{2-5} 
 &
  Tampering &
  \begin{tabular}[c]{@{}l@{}}Short circuit\\ Prevent energy delivery\end{tabular} &
  High &
  \begin{tabular}[c]{@{}l@{}}Anomaly detection\\ Tamper-proof hardware\end{tabular} \\ \cline{2-5} 
 &
  Malicious Code Injection &
  \begin{tabular}[c]{@{}l@{}}Modify BMS response to command\\ Collect sensitive information\end{tabular} &
  Medium &
  \begin{tabular}[c]{@{}l@{}}Authentication\\ Remote attestation\\ Intrusion detection\end{tabular} \\ \cline{2-5} 
\multirow{-10}{*}{\begin{tabular}[c]{@{}c@{}}Battery\\ and\\ BMS\end{tabular}} &
  Spoofing, Replaying, and \acrshort{mitm} &
  \begin{tabular}[c]{@{}l@{}}Report false information to the driver\\ Report false information to the other \acp{ecu}\\ Physically damage the battery\\ Disrupt charging process\\ Excessive discharging\\ Overcharging\end{tabular} &
  High &
  \begin{tabular}[c]{@{}l@{}}Identity management\\ Authentication\\ Intrusion detection\\ Redundancy \\ Timestamps \\Integrity protection \end{tabular} \\ \hline
 &
  \acrshort{mitm} &
  \begin{tabular}[c]{@{}l@{}}Report false information\\ Isolate charger components\\ Modify control signals\\ Increase energy consumption\end{tabular} &
  High &
  \begin{tabular}[c]{@{}l@{}}Anomaly detection\\ Intrusion detection\\ Intrusion prevention\\Integrity Protection\end{tabular} \\ \cline{2-5} 
 &
  \acrshort{dos} &
  Prevent the exchange of energy &
  Medium &
  \begin{tabular}[c]{@{}l@{}}Cookies\\ Time-lock puzzles\\ Rate limiting\\Intrusion detection\end{tabular} \\ \cline{2-5} 
 &
  Spoofing, and Replaying &
  \begin{tabular}[c]{@{}l@{}}Report false information \\ Physical damage\\ Increase energy consumption\end{tabular} &
  High &
  \begin{tabular}[c]{@{}l@{}}Intrusion detection\\ Identity management \\ Timestamps\end{tabular} \\ \cline{2-5} 
 &
  Malicious Code Injection &
  \begin{tabular}[c]{@{}l@{}}Modify \ac{ev} response to commands\\ Collect sensitive information\\ Remote control/hijack\end{tabular} &
  High &
  \begin{tabular}[c]{@{}l@{}}Authentication\\ Remote attestation\\ Intrusion detection\end{tabular} \\ \cline{2-5} 
 &
  Tampering &
  \begin{tabular}[c]{@{}l@{}}Impair the charging process\\ Power loss and overvoltage\end{tabular} &
  High &
  \begin{tabular}[c]{@{}l@{}}Anomaly detection\\ Tamper-proof hardware\end{tabular} \\ \cline{2-5} 
\multirow{-17}{*}{\begin{tabular}[c]{@{}c@{}}Controller\\ and\\ Charger\end{tabular}} &
  Eavesdropping, and Side Channels &
  \begin{tabular}[c]{@{}l@{}}Track the user\\ Profile users' preferences\end{tabular} &
  Low &
  \begin{tabular}[c]{@{}l@{}}Differential privacy\\ Encryption\end{tabular} \\ \hline
\end{tabular}%
}
\end{table*}
} % array stretch closing

\subsection{Battery and BMS}\label{sec:battAndBMS}

The battery pack is a sensitive component of an \ac{ev}. In case of malfunctions, it may catch fire and even explode~\cite{sutcliffe2022look, blum2016fire}. %It is not difficult to imagine how a successful cyberattack can cause such kind of faults, similar to what happens with other \ac{cps} in history~\cite{langner2011stuxnet}. 
Such situations can severely harm the passengers and create financial damage to the owner and a reputation loss to the manufacturer. Less severe cyberattacks can, however, create financial damage, for instance, by reducing the battery's lifespan, forcing the owner to a premature battery replacement~\cite{sripad2019Vulnerabilities}.

The battery pack is managed by the \ac{bms}, which handles communication with the other \acp{ecu} via the vehicle bus. Again, this channel has been proven to be vulnerable to many cyberattacks~\cite{jo2021survey, bozdal2020evaluation}.
In the following, we discuss how cyberattacks impact \acp{ev}, extend their effects to the \ac{cps} domain, and highlight their effect on the battery and \ac{bms}.

\paragraph{Denial of Service} The \ac{bms} is responsible for reporting information on the battery status and managing the energy delivery. An attacker might flood the \ac{bms} controller by forging and sending a vast number of requests, in similar ways to what may happen with \ac{dos} attacks against websites~\cite{gunasekhar2014survey}.
An overload of the \ac{bms} may slow responses to legitimate requests or even prevent the \ac{bms} from sending response messages completely. This may lead to multiple effects depending on the information requested to the \ac{bms} and how the requester device reacts to the absence of a response. In fact, this might cause damage to the battery if power is not properly removed in case of abnormal behavior or physical tampering by the attacker. A \ac{dos} may target sensor measurements, such as temperature, and it may prevent the activation of cooling mechanisms, forcing the battery into critical temperatures, which may be irreversible~\cite{culler2021Cybersecurity}. Furthermore, this attack may also prevent the user from obtaining information on the amount of charge left, causing range anxiety and possibly jeopardizing the drivers' safety in case of a sudden \ac{ev} stop. 

% countermeasure
Flow control might prevent the \ac{bms} from handling many fake requests. In this context, source authentication may provide information regarding the legitimacy of the sender~\cite{groza2018security}. A solution for flow control may be given by an adapted version of time-lock puzzles~\cite{rivest1996time}. Furthermore, rate limiting can help mitigate against \ac{dos} attacks~\cite{kuerban2016flowsec}, while intrusion detection strategies can help in identifying the attack before it creates damage~\cite{young2019survey}. Redundancy on the controllers can also help in mitigating severe \ac{dos} attacks against the \ac{bms}~\cite{bogosyan2020Novel}.

\paragraph{Tampering} An attacker might physically tamper with the battery and the \ac{bms}. Depending on the specific tampered component, an attacker may be able to cause a short circuit that may lead to catastrophic events such as the start of a fire that might harm both the vehicle and the passenger. This consideration holds for battery and \ac{bms}, as they both manage high voltages. Tampering may also lead to less severe consequences, such as the \ac{bms} being unable to communicate with the battery or to deliver the full power to the battery during charging. These attacks may also include detaching or cutting cables. 

% countermeasure
As a possible countermeasure, the battery and \ac{bms} shall include an anomaly detection system to prevent applying a voltage to tampered components and causing the aforementioned damages~\cite{sun2022detection}. Another solution may be the physical protection of these components with tamper-proof hardware. For instance, in case of physical tampering, the battery should be designed so that it cannot receive or deliver power~\cite{gennaro2004algorithmic}. The SAE J2464 standard contains safety measures that can also be effective against tampering~\cite{J2464}.

\paragraph{Malicious Code Injection} The battery pack is managed by the \ac{bms}, which is a piece of hardware with firmware onboard. Attackers may try to reverse engineer the software to discover vulnerabilities and build exploit against them~\cite{kileyReverse}. To patch bugs, \acp{ev}' software may be updated over the air or via the charging cable~\cite{buschlinger2019plug}, thus easing the update process for manufacturers and users. However, this represents a security challenge, as software updates need to access the overall \ac{ev} network~\cite{halder2020secure}. A malicious user may inject malware via software update to gain control of the \ac{bms}~\cite{halder2020secure}. By having partial or complete control over it, the attacker may thus impact the normal functioning of the vehicle. For instance, the malware may prevent the \ac{bms} from requesting energy from the battery, causing a blackout in the \ac{ev}. Contrarily, the \ac{bms} may be forced into requesting more energy than needed to speed up the discharging process. Furthermore, thanks to the malware, the attacker may measure other sensitive information of the driver, which may lead to privacy leakages.

% countermeasures
To prevent code injection and its effects, access to the \ac{ev}'s internal network shall be strictly regulated. Possible solutions include the use of external source authentication. In case of a successful injection, it is fundamental to identify and mitigate its effect. To this aim, remote attestation and its collective extension may be used to validate the in-vehicle components~\cite{ambrosin2020collective}. Furthermore, anomaly and intrusion detection techniques may help identify attacks to the in-vehicle network~\cite{young2019survey}. Injection of malicious updates can be detected by integrity verification on the new software, possibly employing a blockchain~\cite{bere2020blockchain}.

\paragraph{Spoofing, Replaying, and Man-in-the-Middle} An attacker may spoof or modify messages to report to the driver false information on the battery \ac{soc}, thus impairing a safe drive. The attacker may also report incorrect information to the charging infrastructure by impersonating the \ac{bms} or modifying information in the middle of the communication. This may cause the charging process to provoke damage to the battery or the \ac{ev} circuitry. Furthermore, an attacker may report false information to prevent the correct exchange of energy from the battery to the \ac{bms}, for instance, by lowering the current demand and preventing the exchange of a sufficient amount of power from the battery to the \ac{ev}. By requiring excessive power, an attacker can discharge the battery faster than expected in a battery exhaustion attack~\cite{nash2005towards} or may force the battery to overcharge, leading to a massive shortening of the battery lifetime~\cite{sripad2019Vulnerabilities}. Finally, through \ac{mitm}, an attacker may modify the voltage values of the battery pack, leading to over-discharging and consequent battery degradation~\cite{guo2016Mechanism}.

% countermeasures
To prevent these attacks, the battery and \ac{bms} should be given an identity, and all messages shall provide source authentication and integrity protection. The cryptographic material shall be embedded in these devices, with examples in trusted platform modules~\cite{kinney2006trusted} or physical unclonable functions~\cite{gao2020physical}, and shall not be disclosed during communications to prevent \ac{mitm} attacks.
An intrusion detection system can help in identifying ongoing attack~\cite{nash2005towards}.
Redundant controllers can be employed to enhance the resilience of the \ac{bms} against adversarial attacks during charging~\cite{bogosyan2020Novel}.
Kim et al.~\cite{kim2020overview} proposed to employ blockchain to provide authentication and access control in the communication between the \ac{bms} and the other devices inside the \ac{ev}.
Strategies to mitigate the effect of an attacker who has gained direct access to the vehicle's bus have been proposed~\cite{jo2021survey}. Some works have considered peculiar features of \ac{ev} to detect spoofing attacks: Guo et al.~\cite{guo2020cyberattack} proposed a physically-guided machine learning method to detect replay and false data injection attacks on the bus. Their system, tested in a \ac{hil} simulation testbed, could identify the attacks with an accuracy of more than 98\%.
Finally, to prevent replaying attacks, designers can consider the addition of timestamps to packets and signals transmitted~\cite{rughoobur2017lightweight}.

%%%%%%%%%%%%% controller and charger 
\subsection{Controller, Charger, and Electric Motors}\label{sec:cnotrollerandcharger}
The controller and charger are fundamental elements that communicate with the \ac{bms} to exchange power to recharge the battery and to feed the \ac{ev} components with energy. 
The charger communicates with the \ac{evse} to negotiate the parameters of the charge. Moreover, it manages the energy received and forward it to the battery pack according to the \ac{bms} requirements. If bidirectional charging is available to \ac{ev}, the charger may also deliver energy from the \ac{ev} battery to the charging column upon request~\cite{pinto2013bidirectional}.  
The controller manages the energy delivered from the battery to the other components. Some of these components are powered by the battery also in petrol-based vehicles, such as the infotainment system or the lights. Others, such as the electric motors, are instead specific to \ac{ev}. The controller sends energy to them following the driver's input, such as the torque pedal pressure.
This section discusses how an attacker may impair their correct functioning.

% on some papers, attacks are divided into: modification, interference, interception, interruption

\paragraph{Man-in-the-Middle}
Modifying the data in the bus may disrupt the regular operation of the charger since the control signals are usually transmitted through this channel. An attacker may isolate certain charger components (e.g., the load relay), leading to a surge in the DC voltage. These attacks can damage the battery causing degradation in the performance and shortening the lifetime of the battery~\cite{dey2021Real}.
An attacker may also modify the signals managing the electric motors by adding noise or other mutations to the original signal. This attack can damage the correct functionality of the motors and put the driver in dangerous situations~\cite{yang2019vulnerability}.

% countermeasures
Mitigation techniques can be applied using algorithms that can detect the attack in almost real-time by monitoring the physical properties of the vehicle, such as sensor data~\cite{chandwani2020cybersecurity, dey2021Real}. To make the receiver aware of possible \ac{mitm} attacks targeting certain packets, integrity protection mechanisms must be in place. Furthermore, intrusion detection and prevention systems that monitor the data exchanged on the bus can also be implemented to strengthen the defense mechanism~\cite{nash2005towards, valenzuela2012real}.

\paragraph{Denial of Service} The operations handled by the charger and controller heavily rely on the sensors reporting information on the charging status. A malicious user can generate a large number of requests to the sensors reporting data or overload the charger and control modules by flooding them with packets, thus preventing the receipt of legitimate messages. If not properly handled, this attack may cause the controller to stop receiving correct state information, impairing the overall state control.

% countermeasures
Flow control may prevent controllers and motors from handling a large number of fake requests similarly to the \ac{bms}, for instance, by employing an adapted version of time-lock puzzles~\cite{rivest1996time}. Source authentication might be employed to verify the sender's lawfulness~\cite{groza2018security}, while rate limiting can help mitigate against \ac{dos} attacks~\cite{kuerban2016flowsec}. Furthermore, intrusion detection can be adopted to identify ongoing \ac{dos} attacks~\cite{young2019survey}.

\paragraph{Spoofing, and Replaying}
An attacker may spoof sensor identities to create multiple packets with legitimate identifications. By exploiting the same concept, an attacker may also report false information to the charger and controller. Therefore, the controller may take actions based on false data. This may cause damage to the hardware, possibly impairing the whole charging system~\cite{chandwani2020cybersecurity}. False data, if correctly crafted, may also impact the electric motors' functionality. For instance, they could force a stop of the motors by sending false control signals.
Encryption can be a countermeasure to spoofing, preventing a malicious user from freely creating new packets. However, replay attacks can be employed to send correct sensor measurements or actuator updates previously recorded from the bus. 
An attacker may also spoof the information from the infotainment system, acceleration pedal, or other energy-hungry devices in the \ac{ev}. The malicious entity may demand a power amount higher than the truly needed one, thus causing higher energy consumption and shortening the battery's lifespan causing the driver to charge the \ac{ev} frequently. The controller also handles the information regarding acceleration and breaking. An attacker may spoof the related sensors to report false state changes to the electricity supplier. For instance, an attacker may spoof the gas pedal and prevent the receipt of the amount of power needed by the driver to speed up. This may cause safety issues, for instance, when the driver needs to surpass another vehicle.

% countermeasures
As already depicted, encryption can only prevent certain kinds of spoofing attacks, but it is insufficient to mitigate replay attacks. To prevent the latter, a combination of unique identifiers and timestamps can be adopted~\cite{rughoobur2017lightweight}. Identity management may be another fundamental countermeasure against these threats~\cite{groza2018security}. In fact, the controller and charger need to have, by design, access to the identities of all legitimate components. The identification of attacks is possible using intrusion and anomaly detection techniques~\cite{young2019survey}.

\paragraph{Malicious Code Injection} Similarly to the \ac{bms} case, controllers, chargers, and motors also contain software, which may often require updates. However, software updates represent a security challenge since it needs access to the overall \ac{ev} network~\cite{halder2020secure}. A malicious user may force the installation of a malicious software update to gain control of some components of the \ac{ev}'s internal network~\cite{halder2020secure}. The attacker may thus impact the safety of the driver. For instance, the malware may cause the \ac{ev} to respond to the driver commands oppositely (e.g., decelerating while pushing on the gas pedal) and may also propagate to all the \ac{ev}'s components. Furthermore, thanks to the malware, the attacker may measure sensitive information on the driver (e.g., location) or profile the driver. A further threat is due to the implementation of controllers and \acp{ecu} via \ac{fpga}. In this case, an attacker may be able to inject malicious software into the central management system via communication lines by manipulating the \ac{fpga} controller~\cite{chandwani2020cybersecurity}. 

% countermeasures
Countermeasures are similar to the \ac{bms} case. The access to the \ac{ev}'s internal network shall be strictly regulated, for instance, using strong authentication mechanisms. Remote attestation and its collective extension may be used to validate the in-vehicle components~\cite{ambrosin2020collective}, while blockchain can have a role in the verification of new software~\cite{bere2020blockchain}. Finally, intrusion detection techniques may help the identification of ongoing attacks~\cite{young2019survey}.

\paragraph{Tampering} An attacker might physically tamper one of the controllers, charger, or motor components. In this case, for instance, the attacker may prevent the charger from correctly detecting the presence of a power source (either wired or wireless), thus impairing the possibility of charging the vehicle. This is the case for proximity sensors. For instance, the attacker may attach a shield to the pin on the \ac{ev} side such that it cannot correctly communicate with the proximity pilot line. Furthermore, an attacker may tamper with the power converter to degrade its quality, causing power losses or overvoltage.

% countermeasure
To prevent physical tampering, controllers, chargers, and motors may implement anomaly detection frameworks to detect the application of a voltage in non-safe situations and react to the attack~\cite{sun2022detection}. Furthermore, these devices can be designed as tamper-proof, so they will stop functioning in case of tampering~\cite{gennaro2004algorithmic}. Although this may impair the vehicle's functioning, it allows for safeguarding the user's safety.

\paragraph{Eavesdropping, and Side Channels} The current exchanged during the charging process leaks features that can be exploited for user tracking and profiling~\cite{brighente2021tell, evscout}. An attacker may attach a module to the charger and controller to collect the current exchanged during the charging process and extract those features, thanks to the absence of encryption methods. For the same reason, an attacker may also eavesdrop on the information exchanged between the controller, the charger, the motors, and the \ac{bms} to launch the attacks mentioned above. An adversary may also analyze the power exchanged between the controller and the infotainment system to obtain users' sensitive information, such as preferences, habits, and passwords. This attack has been shown in other scenarios~\cite{li2016side}, such as smartphone charging, where users' activities can also be detected in case of encrypted traffic~\cite{conti2015can}. Therefore, it is fundamental to include methods to prevent malicious users from accessing the communication among these entities.

%countermeasure
To guarantee the user's privacy, the current exchanged may be altered via a noisy signal that hides the original signal's features similarly to differential privacy in other contexts~\cite{dwork2008differential}. On the receiver side, the components shall be able to guarantee that the input current does not cause any damage to the circuitry. These solutions may hold for all the involved sources of current. Cryptographic methods may not always represent a viable solution, as they would add computational overhead to a possibly safety-critical system. Furthermore, they do not represent a solution to side channels, which are challenging to mitigate~\cite{khan2014side}.

%%%%%%%%%%%%% EV CHARGING SECURITY %%%%%%%%%%%%%%
%%%%%%%%%%%%%%%%%%%%%%%%%%%%%%%%%%%%%%%%%%%%%%%%%
\section{EV Charging Security and Privacy Challenges}\label{sec:charginSep}
This section discusses the security and privacy issues related to the \ac{ev} charging process. In particular, we discuss the challenges associated with wired charging in Section~\ref{sec:wired_chall}. Then, we discuss the challenges associated with \ac{wpt} and \ac{wipt} in Section~\ref{sec:wptChallenges}. For both technologies, we also discuss possible solutions and countermeasures. In Table~\ref{tab:charginChallenges}, we summarize all the security and privacy challenges associated with the charging process, their effects, impact severeness, and possible countermeasures.

\renewcommand{\arraystretch}{1.2}
\begin{table*}[h!]
\label{tab:charginChallenges}
\caption{Summary of \ac{ev} Charging Challenges.}
\centering
\resizebox{\textwidth}{!}{%
\begin{tabular}{|c|l|l|l|l|}
\hline
\textbf{System} &
  \textbf{Attack} &
  \textbf{Effect} &
  \textbf{Impact} &
  \textbf{Possible Solutions} \\ \hline
 &
  Tampering &
  \begin{tabular}[c]{@{}l@{}}Prevent charging\\ Cause a shock to the driver\\ Get sensitive information\end{tabular} &
  High &
  \begin{tabular}[c]{@{}l@{}}Tamper-proof hardware\\Inconsistencies handler\end{tabular} \\ \cline{2-5} 
 &
  Energy repudiation &
  \begin{tabular}[c]{@{}l@{}}Cheat on billing\\ Steal energy from the system\end{tabular} &
  Low &
  \begin{tabular}[c]{@{}l@{}}Aggregate signature schemes\\ Blockchain for energy transactions\end{tabular} \\ \cline{2-5} 
 &
  \ac{dos} &
  \begin{tabular}[c]{@{}l@{}}Prevent \ac{ev} charging\\ Disruption of the charging service\end{tabular} &
  Medium &
  \begin{tabular}[c]{@{}l@{}}Identity verification\\ Authentication\\Intrusion detection\end{tabular} \\ \cline{2-5} 
 &
  \ac{mitm} &
  \begin{tabular}[c]{@{}l@{}}Prevent proper charging\\ Modify charging parameters\end{tabular} &
  High &
  \begin{tabular}[c]{@{}l@{}}Integrity protection\\ Encryption\end{tabular} \\ \cline{2-5} 
 &
  Spoofing, and Replaying &
  \begin{tabular}[c]{@{}l@{}}Create charging state inconsistencies\\ Steal energy from another \ac{ev}\end{tabular} &
  Medium &
  \begin{tabular}[c]{@{}l@{}}Identity management\\ Authentication\\ Encryption \\ Timestamps \end{tabular} \\ \cline{2-5} 
 &
  Relaying &
  Steal energy from another \ac{ev} &
  Medium &
  \begin{tabular}[c]{@{}l@{}}Distance bounding \\ Fingerprinting\end{tabular} \\ \cline{2-5} 
 &
  Eavesdropping &
  Steal sensitive \ac{ev} information &
  Medium &
  Encryption \\ \cline{2-5} 
\multirow{-15}{*}{Wired Charging} &
  Side Channels, and Information Leaking &
  \begin{tabular}[c]{@{}l@{}}Track user\\ Profile user's preferences\end{tabular} &
  Low &
  \begin{tabular}[c]{@{}l@{}}Differential privacy\\ Secondary batteries\end{tabular} \\ \hline
 &
  Overpower &
  Damage to \ac{ev} battery &
  High &
  \begin{tabular}[c]{@{}l@{}}Energy-efficient overvoltage protection\\Anomaly detection\end{tabular} \\ \cline{2-5} 
 &
  Jamming, and \ac{dos} &
  Prevent \ac{ev} charging &
  Medium &
  \begin{tabular}[c]{@{}l@{}}Channel hopping\\ Identity verification\\ Authentication\\Intrusion detection\end{tabular} \\ \cline{2-5} 
 &
  Freeride attack &
  Steal energy from the system &
  Low &
  \begin{tabular}[c]{@{}l@{}}Authentication\\ Blockchain\end{tabular} \\ \cline{2-5} 
 &
  Energy repudiation &
  \begin{tabular}[c]{@{}l@{}}Cheat on billing\\ Steal energy from the system\end{tabular} &
  Low &
  \begin{tabular}[c]{@{}l@{}}Aggregate signature schemes\\ Blockchain for energy transactions\end{tabular} \\ \cline{2-5} 
 &
  Spoofing, and Replaying &
  \begin{tabular}[c]{@{}l@{}}Create charging state inconsistencies\\ Steal energy from another \ac{ev}\end{tabular} &
  Medium &
  \begin{tabular}[c]{@{}l@{}}Authentication\\ Encryption\\ Physical layer authentication \\ Timestamps\end{tabular} \\ \cline{2-5} 
 &
  Relaying &
  Steal energy from another \ac{ev} &
  High &
  Distance bounding \\ \cline{2-5} 
 &
  \ac{mitm} &
  \begin{tabular}[c]{@{}l@{}}Prevent proper charging\\ Modify charging parameters\end{tabular} &
  Medium &
  \begin{tabular}[c]{@{}l@{}}Integrity protection\\ Encryption\\ Physical layer authentication\end{tabular} \\ \cline{2-5} 
 &
  Eavesdropping &
  Steal sensitive \ac{ev} information &
  Medium &
  Encryption \\ \cline{2-5} 
\multirow{-17}{*}{\ac{wpt}} &
  Side Channels, and Information Leaking &
  \begin{tabular}[c]{@{}l@{}}Track user\\ Profile users' preferences\end{tabular} &
  Medium &
  \begin{tabular}[c]{@{}l@{}}Differential privacy\\ Secondary batteries\end{tabular} \\ \hline
\end{tabular}%
}
\end{table*}

%%%%%%%%%%%%%%%%%%%%%% WIRED %%%%%%%%%%%%%%%%%%%%
\subsection{Wired Charging Challenges}\label{sec:wired_chall}
In the following, we focus on attacks targeting a wired charging scenario, which is the most common way at the moment of writing. Some of the attacks are specific to cases where \ac{hlc} is available (e.g., \ac{mitm}, spoofing), while others are suitable for every type of wired charging, such as tampering or side channel analysis.

\paragraph{Tampering Attacks} In this attack, a malicious user physically tampers with the devices involved in the charging process. In particular, an attacker might manipulate the pilot lines and tamper with the proximity sensor to prevent an \ac{ev} from deeming a secure connection and hence prevent charging. Furthermore, this can also impact users' safety, as it might be possible to detach the cable before removing the current. By observing electromagnetic leaks or operations in the chip components both in the \ac{evse} and \ac{ev}, an attacker might infer sensitive information on the user, such as private keys used for billing purposes~\cite{gottumukkala2019cyber}. Furthermore, by tampering with the charging cable, an attacker might prevent the proper charging of the victim \ac{ev} or steal energy from an \ac{ev} in charge by connecting additional cables~\cite{baker2019Losing, conti2022evexchange}.

%countermeasure
As possible countermeasures, tamper-proof hardware may represent a viable solution~\cite{gennaro2004algorithmic, ahmad2022hardlog}. Thanks to these devices, the attack may be limited to the car functioning without impacting the users' safety. 
%The use of location-based certificates may prevent the attacker from stealing energy from a victim \ac{ev}, although the charging cable may have been modified.
Lastly, proper inconsistency handling mechanisms may be implemented to check that all involved components report the same physical status.

\paragraph{Energy Charging Repudiation} A malicious user may report to the \ac{evse} that the \ac{ev}'s battery did not receive any power by exploiting the behavior of the pilot line and the feedback associated with it. In this situation, the attacker may be able to charge a smaller amount compared to the amount of energy effectively used. If bidirectional charging is available, an attacker may pretend to have sold more energy than it has actually sold, thus stealing money from the energy provider.

% countermeasures
A possible countermeasure to energy repudiation is the use blockchain technology to handle transactions and guarantee traceability and non-repudiation~\cite{jurdak2021trusted, huang2018lnsc}. Aggregate signature schemes from different physical components can represent another possible mitigation to the problem~\cite{zhang2014notes}.

\paragraph{Denial of Charging}
A malicious actor may try to prevent a vehicle from charging. It may be done at the data level by modifying values on the packets exchanged during the handshake between the \ac{ev} and \ac{evse}~\cite{miniv2g}. In some cases, \ac{dos} can also be performed remotely, exploiting unshielded cables, which are often used for the recharge~\cite{kohler2022Brokenwire}.
\ac{dos} may also be launched against more than one vehicle, trying to compromise a portion of the grid.
A greedy attacker may falsify the information on the battery's \ac{soc}, such that s/he can demand an energy amount higher than needed, thus preventing other users from benefiting from the service. The number of users that can simultaneously charge their \acp{ev} and the energy effectively delivered each moment depends on the grid's capacity. 
If the grid capacity is limited, the attacker can successfully launch this attack and prevent other users from charging. 

% countermeasures
Possible countermeasures to \ac{dos} attacks include low-complexity authentication services in all the packets exchanged such that the \ac{evse} can rapidly decide whether to accept or discard a request. Identity-based traffic filtering may be combined with a physical state update related to the charge level of a certain user to prevent multiple malicious requests. Intrusion detection can be employed to detect ongoing \ac{dos} attacks which may generate strange communication patterns~\cite{young2019survey}. Furthermore, enforcing physical security by adopting shielded cables can prevent some kinds of \ac{dos} and eavesdropping attacks~\cite{kohler2022Brokenwire}.

\paragraph{Man-in-the-Middle} When operating charging modes employing \acp{hlc}, such as CHAdeMo or ISO 15118, the \ac{ev} and \ac{evse} exchange data through network packets. A \ac{mitm} attack can be employed to modify the content of this communication. It may be a consequence of tampering if the malicious actor can insert a device on the pilot line between the vehicle and the charging column. In some cases, \ac{mitm} can be performed from other charging columns attacking the SECC Discovery Protocol~\cite{dudek2019v2g}. A malicious actor may exploit this channel to manipulate the exchanged information and create inconsistencies in the recharging process. For instance, an attacker who can modify packets on the fly may prevent proper charging by modifying request and response parameters. Further attacks can be launched starting from \ac{mitm}, such as malware injections or \ac{dos}~\cite{miniv2g}.

% countermeasures 
To identify modified data, integrity protection can be added to packets~\cite{ahmadvand2019taxonomy}. Another possible countermeasure is encryption. Novel versions or ISO 15118 mandates the usage of \ac{tls} for all the communications between the vehicle and charging column, even if in real life, data are often exchanged in plaintext~\cite{baker2019Losing}.

\paragraph{Spoofing, and Replaying}
An attacker might interact in the communication link between the vehicle and the charging column by injecting packets spoofing other devices' identities. For instance, a malicious user can spoof the identity of an \ac{ecu} and report false information on the battery \ac{soc}. Furthermore, an attacker may inject false information by spoofing the identity of an \ac{evse} and stealing sensitive information from an \ac{ev}. For example, in the case of automatic billing based on the \ac{ev} features, a malicious user can extrapolate those features from an \ac{ev} and store them for later use to bill the victim. The same concept can also be applied to other types of connectors, as long as billing is based on automatic feature recognition~\cite{autocharge}.

% countermeasure
Possible countermeasures to these attacks include using a proper identity management scheme, authentication, and data encryption~\cite{chan2015secure}. Authentication systems shall include information related to the charging status of the \ac{ev} or the energy delivered by the \ac{evse} to help guarantee the consistency between the reported information and the actual physical state. It is important to consider that encryption cannot prevent the replaying of packets, which may instead be enforced with unique identifiers and timestamps~\cite{rughoobur2017lightweight}.

\paragraph{Relaying}
A relay attack is possible if an attacker has access to the network traffic and can relay it to a nearby charging column. By relaying information, a malicious user can manipulate the billing system. For instance, a malicious user can relay the data between two neighboring \acp{evse} to bill a closely-located victim user for a charging session~\cite{conti2022evexchange}. If bidirectional charging is available, a malicious user can sell the energy of a victim's vehicle and get paid for it. 

%countermeasures
The location information of \ac{ev} and \ac{evse} may be exploited to prevent relay attacks, e.g., employing distance bounding protocols~\cite{chan2013smart, conti2022evexchange}. Furthermore, the physical features of the \ac{ev} may be exploited to design dedicated authentication protocols~\cite{chan2014cyber, houser2017ev}.

\paragraph{Eavesdropping}
An attacker may be able to read the information exchanged between the vehicle and charging columns in different ways, similarly to what was presented before for \ac{mitm} attacks. With access to all the network traffic, a malicious entity can steal sensitive information from the user, from simple charging parameters to credit card numbers.

%countermeasures
To protect against eavesdropping, encryption can be applied. As already explained, novel versions of ISO 15118 mandate the usage of \ac{tls} for all the communications between the vehicle and the charging column, even if real-life data are still often exchanged in plaintext~\cite{baker2019Losing}. It is important to recall that even if the exchanged data are encrypted, side channel analysis is possible to extract some users' preferences, as presented in the following section.

\paragraph{Side Channels, and Information Leakages} 
An attacker in control of an \ac{evse} may be able to track and profile users who authenticate to the \ac{evse} even if data are encrypted. It may rely on different information, such as the MAC address of the \ac{ev} or the certificate employed by Plug and Charge~\cite{iso15118-1}. However, in Level~1 and Level~2 charging, these kinds of data are unavailable since no \ac{hlc} is generated between the two entities. In that case, an attacker may rely on other features, such as the exact voltage of the control pilot pin or the duration of the handshake at the beginning of the charging process~\cite{houser2017ev}.
Another side channel that may transfer information is the effective current exchange. This does not convey information in a network sense, i.e., it does not involve the creation of packets with the sender's and receiver's information. Therefore, no encryption method is applied to this signal, which is transmitted in plaintext. However, it has been shown that it is possible to profile users by extracting features from the charging current~\cite{brighente2021tell, evscout}. In particular, the charging current contains features peculiar to each \ac{ev}, allowing for \ac{ev} tracking and user profiling based on the current demand. Therefore, it is fundamental to manipulate the current signal to prevent these attacks. 

% countermeasures
Countermeasures to privacy threats shall not undermine the efficiency of the charging process. Therefore, possible solutions must allow the involved parties to retrieve sufficient information, e.g., to the \ac{soc}. Differential privacy methods may represent a viable solution~\cite{hassan2019differential}. An alternative is represented by the use of secondary batteries to create a connection between the \ac{ev} and \ac{evse}, similarly to what was discussed in~\cite{brighente2021tell, evscout}. When \ac{hlc} is available, MAC address randomization may represent a good mitigation technique to reduce the profiling power of an attacker.

%%%%%%%%%%%%%%%%% WPT CHALLENGES %%%%%%%%%%%%%%%%%%%%%%%%%%%%%%%%%%

\subsection{WPT Challenges}\label{sec:wptChallenges}
Due to the exposure of the wireless medium, \ac{wpt} incurs in a large number of safety, security, and privacy issues. In fact, it is likely that \ac{wpt} signals to impact more vehicles and that an attacker gets access to the signals or information wirelessly exchanged~\cite{liu2016safe}. In this section, we review and extend the taxonomy of the possible attacks to \ac{wpt} presented in~\cite{liu2016safe} and adapted it to the \ac{ev} case.

\paragraph{Overpower attack} The wireless medium's intrinsic vulnerability makes it possible that a single \ac{ev} receives both its signal and the signal intended for another vehicle. For instance, if two cars are closely located, and both are charging their batteries via \ac{wpt}, they will receive more power than expected. This is even more likely when considering fully-dynamic \ac{wpt}, where vehicles move and cannot hence guarantee that a reasonable safety space is kept between them. The excessive received power might harm some components of the \ac{bms} or the battery if a proper overvoltage regulator is not deployed. Furthermore, an attacker might exploit this concept to launch an overvoltage attack to damage the \ac{ev}'s components. %Thus, \ac{ev} equipped with \ac{wpt} modules must implement an overvoltage protection mechanism to prevent possible damages.

%countermeasures
Possible countermeasures include implementing overvoltage protection mechanisms at the \ac{ev}'s side. Such mechanisms shall, however, guarantee the efficiency of the charging process to avoid requiring excessive charging times. Anomaly detection methods can also be applied to detect the reception of abnormal power values or other anomalies in the charging process. Design choices can help mitigate overpower attacks. For instance, the distance between coils must be designed to make overpower attack unfeasible or, at least, more complicated.

\paragraph{Jamming, and Denial of Service} In the case of \ac{wipt}, the reception of multiple signals might cause excessive interference at the receiver's side, thus preventing the correct reception of messages. Due to the openness of the \ac{wpt} medium, an attacker might be able to simultaneously jam multiple \acp{ev} by sending random \ac{wipt} messages and degrading the channel quality up to the point where messages are not correctly received. Furthermore, this concept can be exploited to prevent a successful charging negotiation phase, thus preventing a connection between the \ac{ev} and the charging system. This represents a \ac{dos} attack. Similarly, an attacker may launch a jamming attack against the charging column's WiFi access point, preventing legitimate users from connecting and using the service.
An attacker may also target a portion of the energy grid by continuously sending charging requests. If many users engage in this session, they might prevent other users from benefiting from the service availability. Although feedback mechanisms to report on the \ac{soc} of the receiver might be implemented to automatically detach an \ac{ev} when fully charged, an expert attacker might be able to craft feedback packets to avoid showing full battery's \ac{soc}.

% countermeasures
Possible solutions include frequency hopping mechanisms, where channels are selected according to different strategies to avoid using a channel under jamming attack~\cite{grover2014jamming}. Low-complexity authentication services in all the packets exchanged such that the \ac{evse} can rapidly decide whether to accept or discard a request can help in preventing \ac{dos} attacks. To detect a \ac{dos} attack, intrusion detection systems can be deployed~\cite{young2019survey}. Identity-based traffic filtering may be combined with a physical state update related to the charge level of a specific user to prevent multiple malicious requests.

%\paragraph{Attacks to service availability} The number of \acp{ev} that can be simultaneously served depends on the capacity of the grid supplying power to the charging pads. Therefore, resources shall be properly managed to guarantee that all users have access to the charging service. However, greedy users might engage in a unlimited charging session by continuously sending charging requests. If a large number of users engages in this type of sessions, they might prevent other users to benefit from the service availability. Although feedback mechanisms to report on the \ac{soc} of the receiver might be implemented to automatically detach an \ac{ev} when fully charged, an expert attacker might be able to craft feedback packets to avoid showing full battery's \ac{soc}. 

% countermeasures 
%To guarantee the service availability, it is fundamental to ensure the non-malleability of packets reporting \ac{soc} information or to implement a detection mechanism based on historical information to identify malicious users. To this aim, signature methods including physical information regarding the users' battery \ac{soc} may be exploited to guarantee that malicious users do not report false information~\cite{yang2020lis}. \rev{Furthermore, it is important to authenticate users and to prevent the same users to be engaged in more than one charging process. A possible solutions is standardized into ISO 15118~\cite{iso15118-20} and employ certificates to authenticate the users to the system.}

\paragraph{Freeriding attack} As previously mentioned, a user might connect to public infrastructure and pay for charging via \ac{wpt}. Due to the openness of the \ac{wpt} medium, a malicious user could exploit the proximity to a vehicle in charge to steal energy and charge his/her \ac{ev}. A similar scenario envisions the collusion of multiple \ac{ev} owners when a single one registers for the service and multiple users share the bill and benefit from the charging process. These attacks are feasible in all types of dynamic \ac{wpt} models; the only requirement is a short inter-\ac{ev} distance. This attack is challenging to detect, as it does not impact the legitimate channel. In fact, although a second \ac{ev} might be connected to the charging channel, the main channel will not face any performance degradation, thus making it unfeasible to detect the attack. 

% countermeasures
\ac{wpt} sessions need to be authenticated to prevent other users from benefiting from a charging session they are not paying for. Furthermore, authentication procedures might include the physical features of the involved devices and the amount of power transferred. The blockchain solutions proposed by Jiang et al.~\cite{jiang2019blockchain} may be adapted to the \ac{ev} case to guarantee security against this attack.

\paragraph{Energy repudiation} \ac{wpt} is less efficient compared to its wired counterpart, as the wireless medium is characterized by losses due to both attenuation and the relative position of the transmitter and receiver devices. Therefore, part of the transmitted energy may be lost during the charging process. A fair system requires that users pay for the actually received energy. Therefore the billing system needs to compare the transmitted power with the received one. However, this might create security issues. In fact, a malicious user might continuously report a received power value smaller than the true one or report zero received energy. This is commonly known as a repudiation attack, where the user denies benefiting from a service. 

% countermeasures
To guarantee the correctness of the reported power usage information, possible solutions might include the use of aggregate signature schemes from different physical components~\cite{zhang2014notes} or the blockchain technology~\cite{huang2018lnsc, jiang2019blockchain}.

\paragraph{Spoofing, and Replaying} 
A malicious user who knows the standard employed or which is able to eavesdrop on the communication can easily craft malicious packets. Based on the crafted information, this class of attacks may have different impacts on the system. For instance, an attacker may use the identifier of another vehicle to negotiate a charging session that the victim will pay for. Furthermore, a malicious actor can craft packets declaring weird \ac{soc} and spoof other vehicles' identifiers to create inconsistencies in the charging process.

% countermeasures
The use of authentication and integrity protection mechanisms can be effective countermeasures against spoofing. In this context, using physical layer authentication may help in designing suitable protocols~\cite{xie2020survey}. In the context of \ac{wpt}, the transmission frequency can be regulated to encrypt information and guarantee that only the legitimate party can receive power~\cite{zhang2014energy}. This also represents a possible solution to the attacks aforementioned in this section. Finally, timestamps can be added to identify multiple sending of the same packet in a replaying attack~\cite{rughoobur2017lightweight}.

\paragraph{Eavesdropping}
Due to the exposure of the wireless medium, an attacker may easily intercept \ac{wpt} packets. These packets may contain different types of information, such as the vehicle identifier, \ac{soc} information, or billing information. 

% countermeasures
Possible solutions include the use of cryptographic techniques to hide information. The newest release of ISO-15118~\cite{iso15118-20} mandates \ac{tls} on every communication.

\paragraph{Relaying}
An attacker may relay information from a victim vehicle to the access point of the attacker's charging column to steal energy~\cite{conti2022evexchange}. This kind of attack work even if the traffic is encrypted since the data is only relayed and not modified. With respect to the wired counterpart, where the attacker has to tamper in some way with the charging column, a wireless relay attack does not need any hardware modification.

%countermeasures
To protect against relay attacks, a distance bounding protocol can be employed to assess if a malicious entity is relaying the network flow~\cite{conti2022evexchange}.

\paragraph{Man-in-the-Middle attack} With respect to wireless charging, when dealing with \ac{wpt}, the interception and forwarding of communication flow are easier due to the openness of the medium~\cite{hwang2008study}. At the same time, directional jamming can be employed in some cases to prevent the receiver from getting both the original and the modified data.
If the communication flow is unencrypted or the cryptography is weak, an attacker can launch a \ac{mitm} attack to modify on-the-fly packets. For instance, a malicious user might modify the information sent by the victim (i.e., report full \ac{soc}) after establishing a connection with the service provider. To perform such an attack, a malicious entity may set up a fake access point and use it to relay the communication to the legitimate one, gaining the ability to modify packets at will.

%countermeasures
Partial solutions include the previously mentioned solutions, such as encryption, authentication, and integrity protection mechanisms. In the context of \ac{wpt}, the transmission frequency can be regulated to encrypt information and guarantee that only the legitimate party can receive power~\cite{zhang2014energy}. Furthermore, physical layer authentication may enhance the security of the authentication process~\cite{paul2008physical}.

\paragraph{Side Channels, and Information Leaking} Although \ac{wpt} signals might be encrypted or avoid sensitive reporting information on the user, the power signal can be exploited for profiling purposes. This attack has been proven feasible for smartphones, where the \ac{wpt} signal analysis reveals information on the user's activity~\cite{la2021wireless}. This might also be the case for \acp{ev}, where an attacker can infer different types of information. This attack is similar to the profiling performed in the wired case, where it might be possible to track a user and obtain information on her habits and power demands. Preventing this attack represents a challenging task, as it cannot be detected, and data encryption is not sufficient~\cite{saltaformaggio2016eavesdropping}. 

% countermeasures
A possible solution is represented by differential privacy, where data is corrupted with a noisy pattern that might prevent inferring sensitive users' data~\cite{dwork2008differential}. Furthermore, as for the wired counterpart, secondary batteries may prevent the attacker from inferring sensitive users' information.

%%%%%%%%%%%%%%%%%%%%%%%%%%%%%%%%%%%%%%%%%%%%%%%%%%%%%%%%%%%%%
%%%%%%%%%%%%%%%%%%%%%%% FUTURE DIRECTIONS %%%%%%%%%%%%%%%%%%%
%%%%%%%%%%%%%%%%%%%%%%%%%%%%%%%%%%%%%%%%%%%%%%%%%%%%%%%%%%%%%
\section{Future Directions}\label{sec:futureDir}

% summary table
{\renewcommand{\arraystretch}{1.2}
% Please add the following required packages to your document preamble:
% \usepackage{graphicx}
% \usepackage[table,xcdraw]{xcolor}
% If you use beamer only pass "xcolor=table" option, i.e. \documentclass[xcolor=table]{beamer}
\begin{table*}[t]
\centering
\caption{Summary table with attacks and countermeasures for each asset. The first row indicate the assets interested by each attack, while the following rows point out which countermeasure is effective against each attack.\\ \bms: \ac{bms} and battery; \controller: Controllers, charger and motors; \wired: wired charging; and \wpt: \ac{wpt}.}
\label{tab:summary}
\resizebox{\textwidth}{!}{%
\begin{NiceTabular}{|
>{}l |
>{}l |
>{}l |
>{}l |
>{}l |
>{}l |
>{}l |
>{}l |
>{}l |
>{}l |
>{}l |
>{}l |
>{}l |
>{}l |}
\hline
 \diagbox{\large{\textbf{Countermeasure}}}{\large\textbf{Attack}}
% \makecell {\large Countermeasure~\textbackslash~Attack \vspace{4cm}}
&
  \multicolumn{1}{c}{\rot{\textbf{DoS}}} &
  \multicolumn{1}{c}{\rot{\textbf{Tampering}}} &
  \multicolumn{1}{c}{\rot{\textbf{MitM}}} &
  \multicolumn{1}{c}{\rot{\textbf{Replaying}}} &
  \multicolumn{1}{c}{\rot{\textbf{Spoofing}}} &
  \multicolumn{1}{c}{\rot{\textbf{Malware}}} &
  \multicolumn{1}{c}{\rot{\textbf{Overpower}}} &
  \multicolumn{1}{c}{\rot{\textbf{Freeride}}} &
  \multicolumn{1}{c}{\rot{\textbf{Jamming}}} &
  \multicolumn{1}{c}{\rot{\textbf{Repudiation}}} &
  \multicolumn{1}{c}{\rot{\textbf{Eavesdropping}}} &
  \multicolumn{1}{c}{\rot{\textbf{Side-Channels}}} &
  \multicolumn{1}{c}{\rot{\textbf{Relaying}}} \\ \hline
\textbf{Affected assets} &
  \bms \controller \wired \wpt &
  \bms \controller \wired &
  \bms \controller \wired \wpt &
  \bms \controller \wired \wpt &
  \bms \controller \wired \wpt &
  \bms \controller \pad &
  \wpt &
  \wpt &
  \wpt &
  \wired \wpt &
  \bms\controller \wired \wpt &
  \bms\controller \wired \wpt &
  \wired \wpt \\ \hhline{|=|=|=|=|=|=|=|=|=|=|=|=|=|=|}
\textbf{Aggregate signature} &
   &
   &
   &
   &
   &
   &
   &
   &
   &
  \wired \wpt &
   &
   &
   \\ \hline
   
\textbf{Anomaly detection} &
   &
  \bms \controller &
   &
   &
   &
   & \wpt
   &
   &
   & 
   &
   &
   &
   \\ \hline
\textbf{Authentication} &
  \pad \pad \wired \wpt &
   &
   \bms \controller \wired \wpt
   &
    &
  \bms \controller \wired \wpt  &
  \bms \controller  &
   &
  \wpt &
  &
   &
   &
   &
   \\ \hline
\textbf{Blockchain} &
   &
   &
   &
   &
   &
   &
   &
  \wpt &
   &
  \wired \wpt &
   &
   &
   \\ \hline
\textbf{Channel hopping} &
   &
   &
   &
   &
   &
   &
   &
   &
  \wpt &
   &
   &
   &
   \\ \hline
\textbf{Cookies} &
  \bms\controller\pad\pad &
   &
   &
   &
   &
   &
   &
   &
   &
   &
   &
   &
   \\ \hline
\textbf{Differential privacy} &
   &
   &
   &
   &
   &
   &
   &
   &
   &
   &
  \bms\controller &
  \bms\controller \wired \wpt &
   \\ \hline
\textbf{Distance bounding} &
   &
   &
   &
   &
   &
   &
   &
   &
   &
   &
   &
   &
  \wired \wpt \\ \hline
\textbf{Encryption} &
   &
   &
   \bms \controller  \wired \wpt &
   &
  \bms \controller \wired  \wpt &
   &
   &
   &
   &
   &
  \bms\controller \wired \wpt &
   &
   \\ \hline
\textbf{Fingerprinting} &
   &
   &
   &
   &
   &
   &
   &
   &
   &
   &
   &
   &
  \wired \\ \hline
\textbf{Flow control} &
  \bms \controller \pad \pad &
   &
   &
   &
   &
   &
   &
   &
   &
   &
   &
   &
   \\ \hline

\textbf{Intrusion detection/prevention} &
   \bms \controller \wired \wpt
   &
   &
  \bms \controller &
  \bms \controller &
  \bms \controller &
  \bms \controller &
   &
   &
   &
   &
   &
   &
   \\ \hline

\textbf{Identity management} &
   &
   &
  \bms \controller &
  &
  \bms \controller &
   &
   &
   &
   &
   &
   &
   &
   \\ \hline
\textbf{Identity verification} &
  \pad \pad \wired \wpt &
   & \pad \pad \wired \wpt
   & 
   & \pad \pad \wired \wpt
   &
   &
   &
   &
  &
   &
   &
   &
   \\ \hline

\textbf{Inconsistencies handler} &
   &
  \pad \pad \wired &
   &
   &
   &
   &
   &
   &
   &
   &
   &
   &
   \\ \hline

\textbf{Integrity protection} &
   &
   &
  \bms \controller \wired \wpt &
   &
   &
   &
   &
   &
   &
   &
   &
   &
   \\ \hline

\textbf{Overvoltage protection} &
   &
   &
   &
   &
   &
   &
  \wpt &
   &
   &
   &
   &
   &
   \\ \hline
\textbf{Physical layer security} &
   &
   &
  \pad \pad \pad \wpt &
  \pad \pad \pad \wpt &
  \pad \pad \pad \wpt &
   &
   &
   &
   &
   &
   &
   &
   \\ \hline
\textbf{Rate limiting} &
  \bms \controller \pad \pad &
   &
   &
   &
   &
   &
   &
   &
   &
   &
   &
   &
   \\ \hline
\textbf{Redundancy} &
   \bms \pad \pad \pad &
   &
  \bms &
  \bms &
  \bms &
   &
   &
   &
   &
   &
   &
   &
   \\ \hline
\textbf{Remote Attestation} &
   &
   &
   &
   &
   &
  \bms \controller &
   &
   &
   &
   &
   &
   &
   \\ \hline
\textbf{Secondary battery} &
   &
   &
   &
   &
   &
   &
   &
   &
   &
   &
   &
  \pad \pad \wired \wpt &
   \\ \hline
\textbf{Tamper-proof hardware} &
   &
  \bms \controller \wired &
   &
   &
   &
   &
   &
   &
   &
   &
   &
   &
   \\ \hline
\textbf{Time-lock puzzles} &
  \bms \controller \pad \pad &
   &
   &
   &
   &
   &
   &
   &
   &
   &
   &
   &
   \\ \hline
\textbf{Timestamps} &
   &
   &
   &
  \bms \controller \wired \wpt &
   &
   &
   &
   &
   &
   &
   &
   &
   \\ \hline
\end{NiceTabular}%
}
\end{table*}
}%array stretch closing 

Looking at the impact of the different attacks in Tables~\ref{tab:inVeh} and~\ref{tab:charginChallenges}, we can conclude that many security issues related to the cyber-physical nature of \acp{ev} may impact the safety of the driver. We notice that some of the attacks and countermeasures discussed can also be applied to other \ac{ev} assets. However, due to space limitations and to avoid being repetitive, we only discussed those we considered to be the most interesting. Nevertheless, we summarize all the attacks and countermeasures in the comprehensive Table~\ref{tab:summary}. Although many threats concern the charging infrastructure, the most severe in terms of safety are related to the in-vehicle network. In fact, the electric component of \acp{ev} may be tampered with or impaired to electroshock the user. Furthermore, the increasing cyber nature of the \acp{ev}' components leads to challenges regarding the coherency of the information coming from the cyber and the physical worlds. Lastly, the increasing interest in the application of the \ac{wpt} technology to \acp{ev} impose significant challenges that still need to be properly addressed from a \ac{cps} point of view.

Based on our analysis, we foresee the following future directions and needs in the field of \ac{ev} \ac{cps} security.
\begin{itemize}
    \item \acrfull{dos} represents one of the most challenging threats in \acp{ev} security. It is known as difficult to prevent, and almost every component of the \ac{ev} can suffer from it. Compromised internal components can attack other \acs{ecu} to compromise the in-vehicle network, but \ac{dos} attacks can be launched from charging columns to \ac{ev} during charging, or vice-versa. In certain cases, \ac{dos} can have an impact not only on a single vehicle but can compromise \acp{evse} in a certain geographical area. To mitigate this risk, not only do all the vehicle entities need to be associated with an identity, but their allowed flow of information (and hence generated traffic) should depend on the vehicle's physical situation. In fact, it might be that a specific \ac{ecu} needs to send messages at a higher rate when the vehicle is experiencing certain physical stimuli. At the same time, all \acp{ecu} shall be guaranteed a sufficient amount of resources. Therefore, future protections against \ac{dos} for in-vehicle networks should account for the physical factors and the possible impact on the whole electric grid.
    
    \item The potential tampering with the \ac{ev} components might represent a significant threat to the user's safety and may also have repercussions on other elements of the vehicle system. All vehicle components shall be equipped with anomaly detection capabilities or should prevent the application of a voltage or current flow in case of tampering. Possible future solutions might include the collective verification of multiple components to make tampering with a single unit ineffective.
    
    \item The increasing attackers' capabilities impose additional challenges in guaranteeing the cyber-security of \acp{ev}. In fact, an attacker might be able to combine multiple attacks to impair the \ac{ev} functioning. To strengthen the defense mechanisms, it is essential to implement in \acp{ev} frameworks collecting information from multiple sources, combining the cyber and the physical world. For instance, verifying the message integrity might employ data from different sensors and actuators to increase the difficulty of information manipulation. Similarly, intrusion detection techniques might combine network data exchanged through the bus with physical signals from sensors to model better the state of the \ac{ev}. This will also help prevent attacks related to malicious \acp{ecu} controlling actuators for mechanical operations (e.g., steering). Future work should consider the \ac{ev}-specific components such as the battery or the charger as data sources regarding the vehicle's state. For instance, the charge and discharge curves of batteries can be modeled by computers with discrete confidence~\cite{batteryModel1, batteryModel2, batteryModel3}. A simple application of these simulations is a reference to identify packets declaring modified \ac{soc}.
    
    \item One of the strengths of \acp{ev} compared to previous generations of vehicles relies on the software managing all their operations. However, this implies that vehicles are more subject to cybersecurity attacks. Some of these attacks may include malware injection into some of the \ac{ev}'s components. To this aim, collective remote attestation can be used to verify the integrity of all the \ac{ev}'s components and prevent possible safety threats. Remote attestation measures should, however, account for the resource-limited nature of \acp{ev}' components and the time-critical nature of the exchanged information.
    
    \item \ac{wpt} is one of the promising technological solutions to alleviate the range anxiety of drivers fearing not reaching their destination with the available charge. Thanks to the charging while driving paradigm, \acp{ev} can be charged during their operation. However, deploying the required public infrastructure poses many security challenges both to the operators and the users. Some examples include the billing process and the openness of the wireless medium. \ac{wpt} related challenges heavily rely on the cyber-physical nature of the overall infrastructure. Therefore, security solutions in this area should account for the coherency of information from the cyber and the physical domains.
\end{itemize}

%%%%%%%%%%%%%%%%% CONCLUSIONS %%%%%%%%%%%%%%%%%%%%
\section{Conclusion}\label{sec:conc}
The increasing market for \acp{ev} demands an in-depth analysis of \ac{ev} technology's security and privacy challenges. In this paper, we provided an overview of the components of an \ac{ev}, focusing on their characteristic components. We provided the basic information needed to understand how in-vehicle communication networks work and which devices need to communicate with one another. We then discussed how an \ac{ev} battery could be charged via wire and \ac{wpt}. We provided the information needed to understand both technologies and discussed the different implementations. We also provided the security and privacy issues of in-vehicle communications and those related to the charging infrastructure. Focusing on a \ac{cps} perspective, we discussed how different attacks might impact both the user and the system's security and privacy. We then discussed possible countermeasures and proposed some future direction to improve the overall \ac{ev} ecosystem security and privacy. We conclude that the \ac{ev} technology currently presents a large attack surface that users with malicious intents can exploit. Therefore, it is fundamental to develop technologies considering the \ac{cps} nature of \acp{ev} to provide full security.

\bibliographystyle{IEEEtran}
\bibliography{biblio}
\vskip 0pt plus -1fil
\begin{IEEEbiography}
[{\includegraphics[width=1in,height=1.15in,clip]{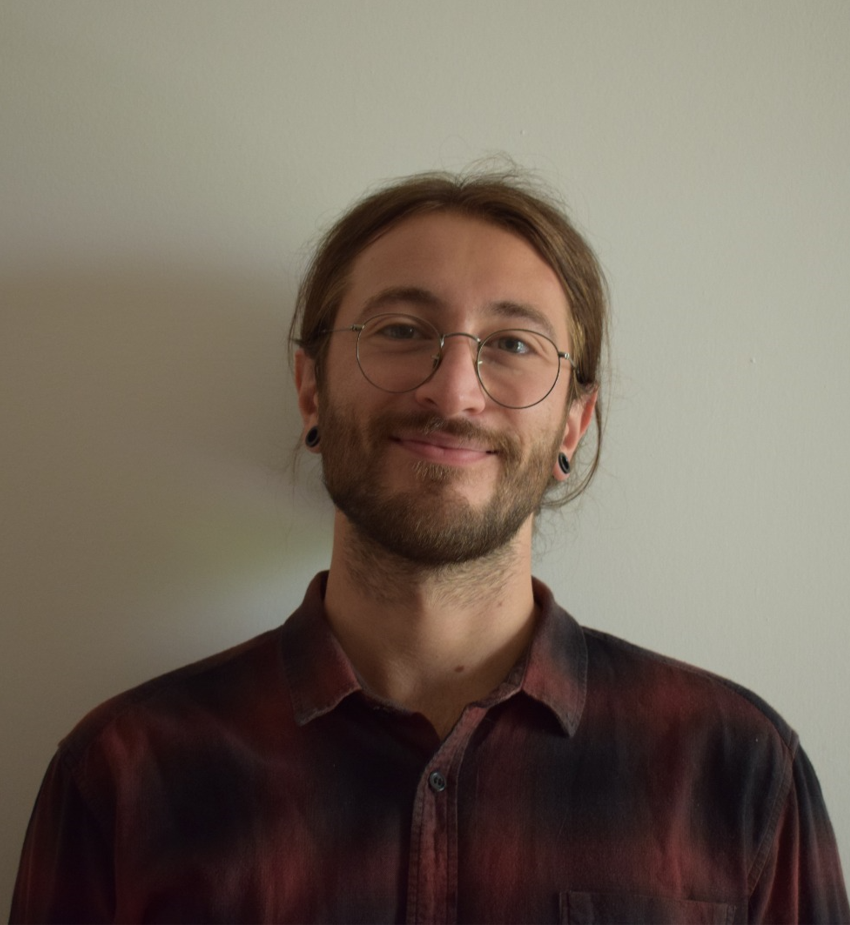}}]%
{Alessandro Brighente} is assistant professor at the University of Padova. He was visiting researcher at Nokia Bell Labs, Stuttgart, Germany in 2019 and University of Washington, Seattle, USA, in 2022. He served as TPC for several conferences, including Globecom, VTC, and WWW. He is guest editor for IEEE Transactions on Industrial Informatics and program chair of DevSecOpsRA, co-located with EuroS\&P. His current research interests include security and privacy in cyber-physical systems, vehicular networks, blockchain, and communication systems.
\end{IEEEbiography}
\vskip 0pt plus -1fil
\begin{IEEEbiography}[{\includegraphics[width=1in,height=1.25in,clip,keepaspectratio]{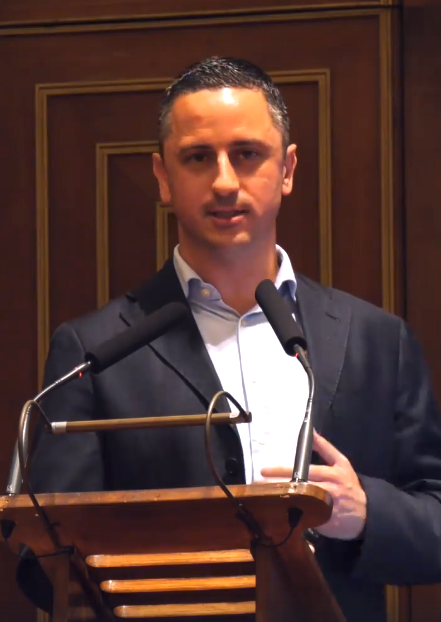}}]{Mauro Conti} is Full Professor at the University of Padua, Italy. He is also affiliated with TU Delft and University of Washington, Seattle. He obtained his Ph.D. from Sapienza University of Rome, Italy, in 2009. After his Ph.D., he was a Post-Doc Researcher at Vrije Universiteit Amsterdam, The Netherlands. In 2011 he joined as Assistant Professor the University of Padua, where he became Associate Professor in 2015, and Full Professor in 2018. He has been Visiting Researcher at GMU, UCLA, UCI, TU Darmstadt, UF, and FIU. He has been awarded with a Marie Curie Fellowship (2012) by the European Commission, and with a Fellowship by the German DAAD (2013). His research is also funded by companies, including Cisco, Intel, and Huawei. His main research interest is in the area of Security and Privacy. In this area, he published more than 400 papers in topmost international peer-reviewed journals and conferences. He is Area Editor-in-Chief for IEEE Communications Surveys \& Tutorials, and has been Associate Editor for several journals, including IEEE Communications Surveys \& Tutorials, IEEE Transactions on Dependable and Secure Computing, IEEE Transactions on Information Forensics and Security, and IEEE Transactions on Network and Service Management. He was Program Chair for TRUST 2015, ICISS 2016, WiSec 2017, ACNS 2020, and General Chair for SecureComm 2012, SACMAT 2013, CANS 2021, and ACNS 2022. He is Senior Member of the IEEE and ACM. He is a member of the Blockchain Expert Panel of the Italian Government. He is Fellow of the Young Academy of Europe.
\end{IEEEbiography}
\vskip 0pt plus -1fil
\begin{IEEEbiography}[{\includegraphics[width=1in,height=1.25in,clip,keepaspectratio]{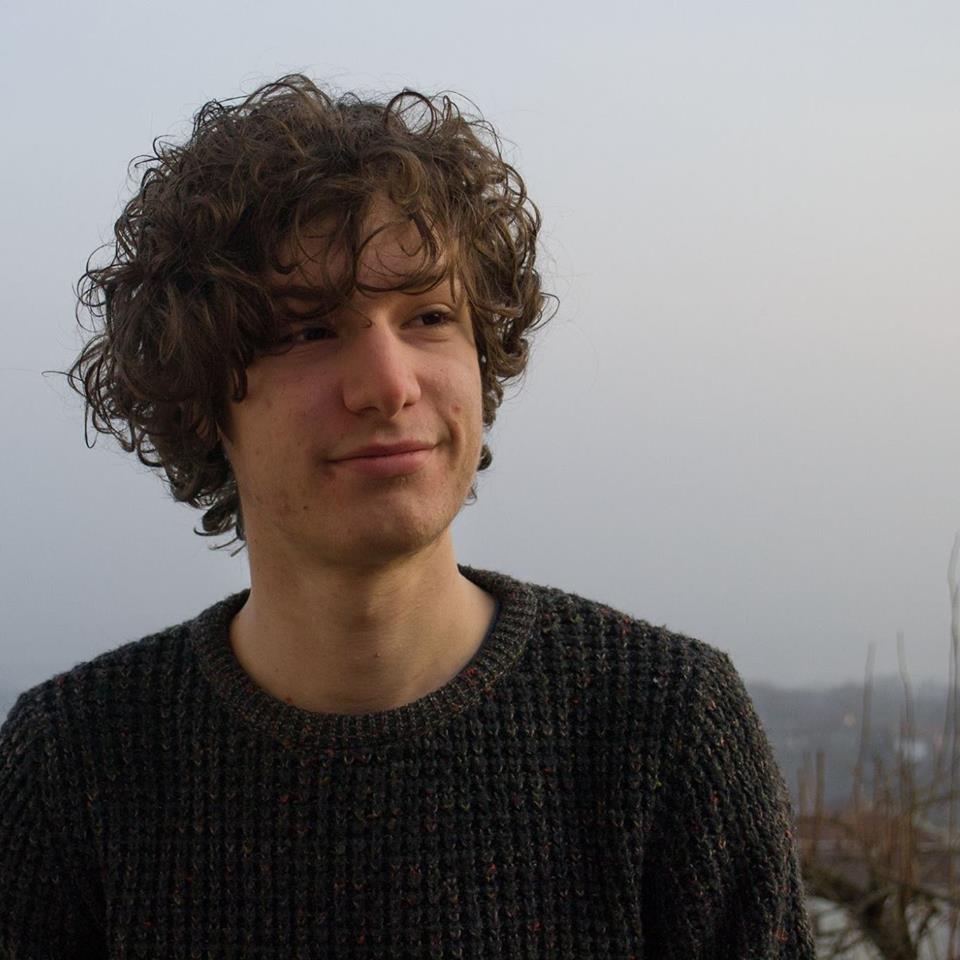}}]{Denis Donadel} received his MSc in Telecommunication Engineering from the University of Padua, Italy, in 2020. He is now a Ph.D. Student in Brain, Mind and Computer Science (BMCS) at the University of Padua where he joined the SPRITZ Security and Privacy Research Group under the supervision of Prof. Mauro Conti. Together with his academic course, Denis is also working with Omitech SRL as part of his high apprenticeship program. During the 2021 Summer, he was granted the New Generation Internet (NGI) Explorers grant to support a collaboration with the University of Washington (Seattle, USA). His research interests lie primarily in Cyber-Physical Systems security, focusing particularly on Vehicles Security and Critical Infrastructures Security.
\end{IEEEbiography}
\vskip 0pt plus -1fil
\begin{IEEEbiography}[{\includegraphics[width=1in,height=1.25in,clip,keepaspectratio]{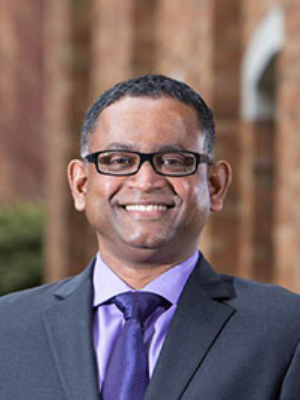}}]{Radha Poovendran}
is Professor of the Department of Electrical \& Computer Engineering at the University of Washington. He is the founding director of the Network Security Lab and is a founding member and associate director of research for the UW's Center for Excellence in Information Assurance Research and Education. He has also been a member of the advisory boards for Information Security Education and Networking Education Outreach at UW. In collaboration with NSF, he served as the chair and principal investigator for a Visioning Workshop on Smart and Connected Communities Research and Education in 2016. 
Poovendran's research focuses on wireless and sensor network security, adversarial modeling, privacy and anonymity in public wireless networks and cyber-physical systems security. He co-authored a book titled Submodularity in Dynamics and Control of Networked Systems and co-edited a book titled Secure Localization and Time Synchronization in Wireless Ad Hoc and Sensor Networks. Poovendran is a Fellow of IEEE and has received various awards including Distinguished Alumni Award, ECE Department, University of Maryland, College Park, 2016; NSA LUCITE Rising Star 1999; NSF CAREER 2001; ARO YIP 2002; ONR YIP 2004; PECASE 2005; and Kavli Fellow of the National Academy of Sciences 2007.
\end{IEEEbiography}
\vskip 0pt plus -1fil
\begin{IEEEbiography}[{\includegraphics[width=1in,height=1.25in,clip,keepaspectratio]{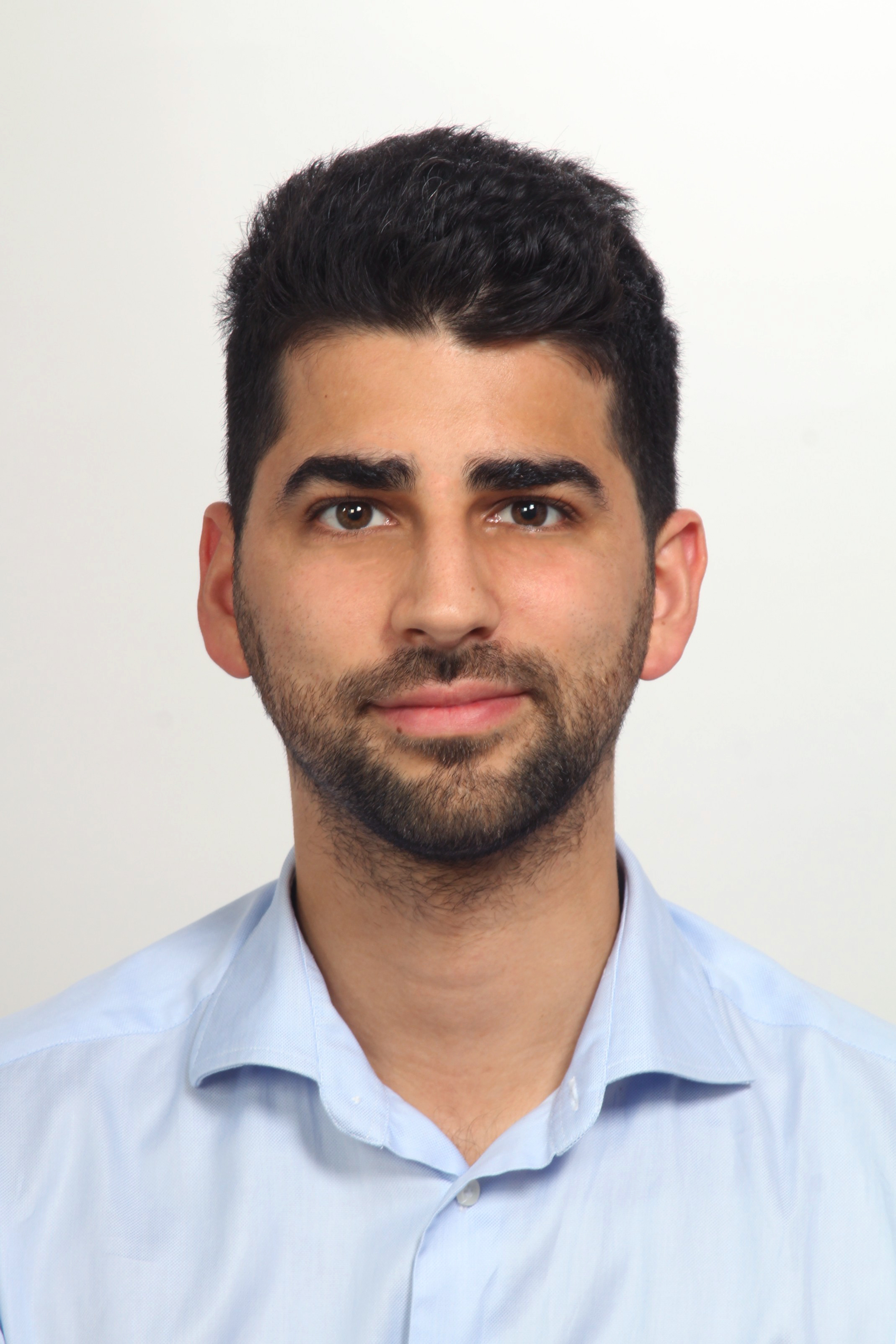}}]{Federico Turrin}
received the Master’s Degree in Computer Engineering from the University of Padova, Italy, in 2019, where he is currently pursuing the interdisciplinary Ph.D. in Brain, Mind, and Computer science, since October 2019. He has been visiting researcher at SUTD Singapore in 2022. His research interests lie primarily in Cyber-Physical System Security with a particular focus on Industrial Control Systems Security, Vehicles Security, and Anomaly detection.
\end{IEEEbiography}
\vskip 0pt plus -1fil
\begin{IEEEbiography}[{\includegraphics[width=1in,height=1.25in,clip,keepaspectratio]{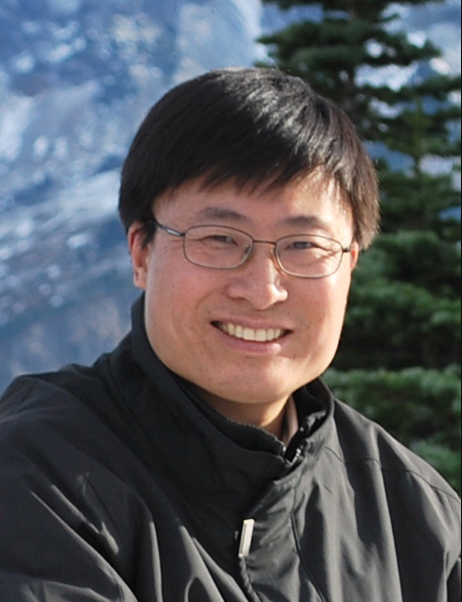}}]{Jianying Zhou}  is a professor and co-center director for iTrust at Singapore University of Technology and Design (SUTD). He received PhD in Information Security from Royal Holloway, University of London. His research interests are in applied cryptography and network security, cyber-physical system security, mobile and wireless security. He has published 300 referred papers at international conferences and journals with 13,000 citations, and received ESORICS'15 best paper award. He has 2 technologies being standardized in ISO/IEC 29192-4 and ISO/IEC 20009-4, respectively. He is a co-founder \& steering committee co-chair of ACNS. He is also steering committee chair of ACM AsiaCCS, and steering committee member of Asiacrypt. He has served 200 times in international cyber security conference committees (ACM CCS \& AsiaCCS, IEEE CSF, ESORICS, RAID, ACNS, Asiacrypt, FC, PKC etc.) as general chair, program chair, and PC member. He has also been in the editorial board of top cyber security journals including IEEE Security \& Privacy, IEEE TDSC, IEEE TIFS, Computers \& Security. He is an ACM Distinguished Member. He received the ESORICS Outstanding Contribution Award in 2020, in recognition of contributions to the community.
\end{IEEEbiography}

\end{document}